%% file: FofRiem-v2.tex
\documentclass[notitlepage, 
11pt, amsmath, preprintnumbers, nofootinbib, aps, floatfix] {revtex4-1}

\usepackage{amssymb,url,bm}
\usepackage{amsfonts,multirow}
\usepackage{pifont}
\usepackage{amsmath}
\usepackage{graphicx}
\usepackage{hyperref}
\usepackage{color}
\usepackage{colortbl}
\usepackage[table]{xcolor}
\usepackage{array,mathtools,amssymb,booktabs}
\usepackage{float}
\usepackage{setspace}
\usepackage{mathrsfs}

\input{layout}

\input{commands}

\begin{document}
\def\s0#1#2{\mbox{\small{$ \frac{#1}{#2} $}}}
\def\0#1#2{\frac{#1}{#2}}

\aboverulesep = 0mm
\belowrulesep = 0mm
\title{Functional Renormalisation  for   $f(R_{\mu\nu\rho\sigma})$ Quantum Gravity}
\author{Yannick Kluth}
\author{Daniel F. Litim}
\affiliation{\mbox{Department of Physics and Astronomy, University of  Sussex, Brighton, BN1 9QH, U.K.}}
\begin{abstract}
We derive    new functional renormalisation group flows for     quantum gravity, in any dimension.
The key new  achievement is that the  equations apply  for {\it any}  theory of gravity whose underlying Lagrangian 
$\sim f(R_{\mu\nu\rho\sigma})$ is a function of  the Riemann  tensor  and the inverse metric. 
The results centrally exploit the benefits of maximally symmetric spaces for the evaluation of operator traces.
The framework is highly versatile and offers a wide  range of new applications to study
quantum gravitational effects in  extensions of Einstein gravity, many of which have hitherto been out of reach. 
The phase diagram and sample  flows  for Einstein-Hilbert gravity,  Gauss-Bonnet,  and selected higher-order theories of gravity are given.
We  also provide  an algorithm to find the flow  for general polynomial Riemann curvature interactions. 
The setup vastly enhances the reach of fixed point searches, enabling   novel types of search strategies 
including across the operator space spanned by polynomial curvature invariants, and  in  extensions of general relativity relevant for cosmology. 
Further implications, and links with unimodular versions of gravity are indicated.

\end{abstract}

\newpage

\maketitle
\begin{spacing}{.95}
\tableofcontents
\end{spacing}
\renewcommand{\baselinestretch}{.8}
\setlength{\parskip}{7pt}

\section{\bf Introduction}\label{sec:intro}

An appealing scenario for the quantum nature of gravity is that general relativity emerges as a relevant perturbation from an interacting UV conformal fixed point \cite{Weinberg:1980gg}. If so, a finite number of independent parameters would ensure predictivity, and characterise the  renormalisation group trajectories which connect the  fixed point of quantum gravity in the UV with classical gravity in the IR. However, what complicates matters at this point is that the fundamental gravitational Lagrangian is not known. In principle, it should consist of an infinite tower of 
interactions formed out of the Riemann tensor and covariant derivatives thereof, where, unlike in BSM model building, higher-dimensional interactions  cannot be omitted. Hence, identifying fixed points and relevant perturbations in quantum gravity would  seem like an impossible task.

Progress has been made in studying subsets of curvature invariants. 
In Einstein-Hilbert gravity,  asymptotically safe fixed points and  relevant  perturbations have by now been identified \cite{Reuter:2001ag,Souma:1999at, Souma:2000vs,
 Lauscher:2001ya,
  Litim:2003vp,
  Bonanno:2004sy,
  Fischer:2006fz,
  Litim:2008tt,   
   Eichhorn:2009ah,
  Manrique:2009uh,
   Eichhorn:2010tb,
   Manrique:2010am,
   Manrique:2011jc,
   Litim:2012vz,
   Donkin:2012ud,
   Christiansen:2012rx,
   Codello:2013fpa,
   Christiansen:2014raa,
   Becker:2014qya,
   Falls:2014zba,
   Falls:2015qga,
   Falls:2015cta,
   Christiansen:2015rva,
   Gies:2015tca,
   Benedetti:2015zsw,
   Biemans:2016rvp,
   Pagani:2016dof,
   Falls:2017cze,
   Houthoff:2017oam,
   Knorr:2017fus,
   Baldazzi:2021orb}.
Next natural steps have been the inclusion of higher-order interactions such as  fourth order interactions, non-local interactions, powers of the Ricci scalar $R$, or  functions of curvature invariants such as $f(R)$ and similar
\cite{
 Lauscher:2002sq,
   Codello:2006in,
   Codello:2007bd,
   Machado:2007ea,
   Codello:2008vh,
 Benedetti:2009rx,
 Benedetti:2009gn,
 Benedetti:2010nr,
   Niedermaier:2011zz,
  Niedermaier:2009zz,
  Niedermaier:2010zz,
 Groh:2011vn,
   Benedetti:2012dx,
   Dietz:2012ic,
Falls:2013bv,
   Ohta:2013uca,
  Benedetti:2013jk,
  Dietz:2013sba,
   Falls:2014tra,
 Saltas:2014cta,
 Demmel:2014sga,
    Eichhorn:2015bna,
   Ohta:2015efa,
   Ohta:2015fcu,
 Demmel:2015oqa,
 Falls:2016wsa,
   Falls:2016msz,
   Gies:2016con,
   Christiansen:2016sjn,
   Gonzalez-Martin:2017gza,
   Becker:2017tcx,
   Falls:2017lst,
 Falls:2018ylp,
   deBrito:2018jxt,
   Knorr:2021slg,
   Falls:2020qhj,
   Kluth:2020bdv,
   Sen:2021ffc,
   Mitchell:2021qjr}
An important tool in the search for fixed points is the bootstrap search strategy  \cite{Falls:2013bv}, where the canonical mass dimension of invariants is used as an ordering principle. High-order bootstrap studies  have shown that quantum gravity becomes ``as Gaussian as it gets'',  and largely dominated by  a few dominant interactions \cite{Falls:2013bv,Falls:2014tra,   Falls:2017lst,Falls:2018ylp,Kluth:2020bdv}.

More concretely, fixed point searches have addressed fourth-order interactions 
\cite{Benedetti:2009gn,Niedermaier:2011zz,  Niedermaier:2009zz,  Niedermaier:2010zz,Falls:2020qhj}
  and a selection of sixth-order interactions including 
$R^3$ \cite{Codello:2007bd,Machado:2007ea,Codello:2008vh,Falls:2013bv,Falls:2014tra},  
the Goroff-Sagnotti term 
 $C_{\mu\nu}{}^{\rho\sigma} C_{\rho\sigma}{}^{\tau\lambda} C_{\tau\lambda}{}^{\mu\nu}$ \cite{Gies:2016con},
 $R \cdot R_{\mu\nu}R^{\mu\nu}$ \cite{Falls:2017lst}, 
and $R\cdot (R_{\mu\nu\rho\sigma}R^{\mu\nu\rho\sigma})$ \cite{Kluth:2020bdv}. 
 Dedicated high-order studies have retained  selected types of  curvature interactions such as
 $R^{2+n}$ \cite{Falls:2013bv,Falls:2014tra,Falls:2018ylp}, 
 $ (R_{\mu\nu}R^{\mu\nu})^n$ \cite{Falls:2017lst}, $R\cdot (R_{\mu\nu}R^{\mu\nu})^{n}$ \cite{Falls:2017lst},  
 $ (R_{\mu\nu\rho\sigma}R^{\mu\nu\rho\sigma})^n$  \cite{Kluth:2020bdv}, and $R\cdot (R_{\mu\nu\rho\sigma}R^{\mu\nu\rho\sigma})^{n}$ \cite{Kluth:2020bdv}, often up to including very high orders $n$. By and large, fixed points and relevant perturbations with viable scaling dimensions are found. Results converge well with increasing number of interaction monomials. Higher curvature interactions are important quantitatively even if they relate, ultimately, to irrelevant perturbations.   More recently, it was noted    that more complex Ricci and Riemann tensor interactions may  shift fixed points  more strongly away from their Einstein-Hilbert counterpart  \cite{Falls:2017lst,Kluth:2020bdv}, and that Riemann tensor interactions may lead to new relevant perturbations in the UV \cite{Kluth:2020bdv}.

These findings encourage broader, systematic investigations. 
In this work, we  provide new functional flow equations  for $f(R_{\mu\nu\rho\sigma})$ quantum gravity.
Crucially, to enable the study of general curvature invariants we take full advantage of maximally symmetric backgrounds which are used for the evaluation of operator traces.
The setup encompasses essentially all gravitational flows investigated thus far within the asymptotic safety programme, with the exception of flows which adopt $e.g.$~less symmetric background geometries.  
We  discuss the pros and cons of our approach, and explain in detail why it leads to important simplifications. 
Most notably, we find that a general flow for $f(R_{\mu\nu\rho\sigma})$ quantum gravity can be determined 
{\it without}  specifying the underlying Lagrangian, other than being of the $f(R_{\mu\nu\rho\sigma})$ type. 
This underlying structure allows for new types of systematic fixed point searches in the space of polynomial curvature invariants,
and  opens the door for much farther reaching fixed points searches beyond.

The remainder of this work is organised as follows. 
In \sct{sct:funcrenorm}, we recall the formalism of  functional renormalisation for gravity within the  single-field or background field formalism,  discuss salient technical aspects and  detail the choices adopted in this work.
In \sct{sct:Bueno}, we derive the renormalisation group  flows for  $f(R_{\mu\nu\rho\sigma})$ theories of gravity, providing all steps of the derivation. 
We explain why the flows on maximally symmetric backgrounds can be parametrised by a small set of scalar functions. 
Further aspects of the methodology, the Hessians, the  flows, and their key new features  are also discussed.
In~\sct{sct:compotherflow}, we illustrate the methodology with several applications, including a general algorithm to find flows for polynomial curvature invariants, new flows and analytical phase diagrams for Einstein-Hilbert gravity, and sample flows for Gauss-Bonnet  and various  higher-curvature gravities.
In \sct{sec:conclusions} we conclude with a brief outlook.
Three appendices summarise technicalities such as metric derivatives of the Lagrangian (App.\ref{sct:metricderiv}), Hessians obtained without the York decomposition (App.\ref{app:Hessian}), and explicit expressions for the gravitational renormalisation group equations (App.\ref{sec:floweq}).

\section{\bf Renormalisation Group for Gravity}
\label{sct:funcrenorm}
In this work, we employ renormalisation group methods to find flow equations of quantum gravitational systems. 
In general, we are interested in non-perturbative effects of quantum gravity for which the functional renormalisation group serves as a useful semi-analytical tool. 
In this section, we give a brief introduction to the functional renormalisation group and explain the  technical setup and key choices made in this work.

\subsection{Functional Renormalisation}
Our starting point is the partition function of a quantum field theory containing quantum fields $\phi_i$, where the index $i$ labels different fields as well as Lorentz indices. For the particular case of pure quantum gravity, $\phi_i$ might contain the metric field $g_{\mu \nu}$ as well as ghost fields arising from the gauge fixing procedure or the measure of the path integral. We denote the classical action of this theory by $S [\phi_i]$. To give the partition function a renormalisation scale dependence an infrared regulator $\Delta S_k$ can be included,
\begin{equation}
	Z_k \left[ j^i \right] = \int \mathcal{D} \phi_i \text{exp} \left\{ \Delta S_k [\phi_i] - S [\phi_i] - \int \text{d}^d x \sqrt{\overline{g}} \, \phi_i j^i \right\} \, ,
\end{equation}
with
\begin{equation}
	\Delta S_k [\phi_i] = \int \text{d}^d x \sqrt{g} \, \phi_i \mathcal{R}^{\phi_i \phi_j}_k (\Delta) \phi_j \, ,
\end{equation}
and $j^i$ the source terms for the fields $\phi_i$. The infrared regulator $\mathcal{R}^{\phi_i \phi_j}_k (\Delta)$ is chosen such that it suppresses modes with $\Delta \phi_j < k^2 \phi_j$, where $\Delta$ is a Laplacian for the field $\phi_i$, and $k$ denotes the RG scale.
Starting from these definitions, the effective average action $\Gamma_k$ is related to the Legendre transformation of the partition function, defined by
\begin{equation}
	\Gamma_k [\phi_i] \equiv \tGamma_k [\phi_i] - \Delta S_k [\phi_i] \, ,
\end{equation}
with
\begin{equation}
	\tGamma_k [\phi_i] =  - \log (Z_k [j_{\phi_j}^i]) + \int \text{d}^d x \sqrt{g} j_\phi^i \phi_i \, , \qquad \fderiv{\tGamma_k}{\phi_i} = j_{\phi_j}^i \, .
\end{equation}
In the infrared limit where the cutoff is removed, $\Delta S_k\to 0$ for $k \rightarrow 0$, the effective action $\Gamma_k$ reduces to the quantum effective action $\Gamma$. The scale dependence of $\Gamma_k$ can be shown to be given by an exact functional identity, the flow equation, which derives from the path integral integral representation of the theory \cite{Wetterich:1992yh},
\begin{equation}
	\partial_t \Gamma_k = \frac{1}{2} \text{Tr} \left\{ \frac{\partial_t \mathcal{R}_k}{\Gamma_k^{(2)} + \mathcal{R}_k} \right\} \, ,
	\label{eqn:wetterich}
\end{equation}
where the only objects entering the right-hand side are the infrared regulator $\mathcal{R}_k$ and the Hessians of the effective average action $\Gamma_k^{(2)}$. The trace on the right-hand side of \eq{eqn:wetterich} is a functional trace including a sum over all field indices as well as an integration over spacetime.

At weak coupling, iterative solutions of the flow generate the conventional perturbative loop expansion \cite{Litim:2001ky,Litim:2002xm}. In the limit where the momentum cutoff becomes a momentum-independent mass term the flow reduces to a Callan-Symanzik equation \cite{Litim:1998nf}, which may require an additional renormalisation of the flow itself~\cite{Fehre:2021eob}. 
The usefulness of \eq{eqn:wetterich} stems from the fact that it is an exact equation,  and that it can be applied in situations where perturbative treatments are no longer applicable. Further, it can be solved exactly in special  limits such as  large-$N$, e.g.~\cite{Tetradis:1995br,DAttanasio:1997yph,Litim:2011bf,Marchais:2017jqc,Litim:2018pxe}. Elsewise practical solutions often involve systematic approximations such as the derivative expansion, vertex expansion, or combinations thereof, giving access to non-perturbative effects. 
Further, optimised choices for the regulator  \cite{Litim:2000ci,Litim:2001up} allow for analytic flows and enhanced convergence \cite{Litim:2001fd,Litim:2005us}. 
The stability of approximations can  be probed through the variation of technical parameters such as the cutoff function  ${\cal R}_k$ \cite{Litim:1996nw,Freire:2000sx,Litim:2001dt,Litim:2001fd,Fischer:2006fz} or the projection method \cite{Litim:2010tt}.

In gravity, the functional renormalisation group has been introduced originally in \cite{Reuter:1996cp}, with many subsequent studies 
testing the asymptotic safety scenario. Analytical flows for gravity have first been provided in \cite{Litim:2003vp} and used in  many  studies of gravity. Further, maximally symmetric backgrounds have been used prominently to evaluate operator traces for gravitational flows  \cite{Codello:2008vh,Falls:2018ylp,Falls:2017lst,Falls:2014tra}. In the following, we explain some of the technical choices used there, and generalise gravitational flows for general actions.

\subsection{Splitting the Metric}
\label{sct:metricsplit}
In quantum gravity, the introduction of an IR regulator requires the usage of the background field method due to the necessity of a scale separating IR from UV modes \cite{Freire:2000bq}. Hence, in this case the full metric needs to be split into a background metric $\bg_{\mu \nu}$ and a fluctuation part $\delta g_{\mu \nu} [h_{\mu \nu}]$ yielding
\begin{equation}
	g_{\mu \nu} = \overline{g}_{\mu \nu} + \delta g_{\mu \nu} [h_{\rho \sigma}] \, .
	\label{eqn:genmetricsplit}
\end{equation}
The path integral then only integrates over the fluctuation field $h_{\mu \nu}$. Note that the fluctuation part can in general depend non-linearly on the fluctuation field. Two natural choices arise, the linear split
\begin{equation}
	g_{\mu \nu} = \bg_{\mu \nu} + h_{\mu \nu} \, ,
	\label{eqn:metriclinear}
\end{equation}
and the exponential split, originally introduced for the study of quantum gravity in  $2+\eps$ dimensions \cite{Kawai:1992np,Kawai:1993mb,Kawai:1995ju,Aida:1994zc} (see also \cite{Nink:2014yya}) 
\begin{equation}
	g_{\mu \nu} = \bg_{\mu \rho} (e^h)^\rho_{\ \nu} = \bg_{\mu \nu} + h_{\mu \nu} + \frac{1}{2} h_{\mu \rho} h^\rho_{\ \nu} + \mathcal{O} \left( h^3 \right) \, .
	\label{eqn:metricexp}
\end{equation}
Using the linear split \eq{eqn:metriclinear} corresponds to an integration over all possible metric fields in the path integral, including degenerate metrics and metrics with different signature than the background metric $\bg_{\mu \nu}$. In contrast to that, the exponential split formally restricts the integration to be carried out over those metrics $g_{\mu \nu}$ which have the same signature as $\bg_{\mu \nu}$.\footnote{For other variants of an exponential split, see \cite{Falls:2015qga,Falls:2015cta}.} 
To be able to study both of these choices simultaneously, we introduce a parameter $\tau$ interpolating between them \cite{Gies:2015tca},
\begin{equation}
	g_{\mu \nu} = (1 - \tau) \left[ \bg_{\mu \nu} + h_{\mu \nu} \right] + \tau \bg_{\mu \rho} ( e^h )^\rho_{\ \nu} = \bg_{\mu \nu} + h_{\mu \nu} + \frac{\tau}{2} h_{\mu \rho} h^\rho_{\ \nu} + \mathcal{O} \left( h^3 \right)\, .
	\label{eqn:taumetricsplit}
\end{equation}
Clearly, for $\tau = 0$ we get \eq{eqn:metriclinear} and for $\tau = 1$ we have \eq{eqn:metricexp}. We will, however, not limit 
ourselves to such choices and implement $\tau$ as a free parameter in our setup. 

As a last remark, we note that details of the metric split \eq{eqn:genmetricsplit}  beyond quadratic order \eq{eqn:taumetricsplit} will not be needed in the present work. This implies that \eq{eqn:taumetricsplit} with a free parameter $\tau$ captures already the most general case. In this light, it  should also be kept in mind that the value  $\tau=1$ (to which we will refer  as the exponential split) can be achieved by many other non-linear splits \eq{eqn:genmetricsplit}  different from \eq{eqn:metricexp},  some of which may formally correspond to metrics $g_{\mu \nu}$ which have a different signature from $\bg_{\mu \nu}$.

\subsection{Background Field Approximation}
Using the background field method with a metric split as in \eq{eqn:genmetricsplit}, the classical action $S [\phi_i]$ only depends on the full metric $g_{\mu \nu}$ and, therefore, its dependence on $\bg_{\mu \nu}$ and $h_{\mu \nu}$ is related by \eq{eqn:genmetricsplit}. This is, however, not true for the regulator term $\Delta S_k [\phi_i]$, since
\begin{equation}
	\mathcal{R}_k^{\phi_i \phi_j} = \mathcal{R}_k^{\phi_i \phi_j} [\bg_{\mu \nu}]
\end{equation}
is only allowed to depend on the background metric $\bg_{\mu \nu}$ but not on the quantum fields, or else the flow equation is no longer exact in its present form.
Hence, the regulator required to obtain \eq{eqn:wetterich} generally breaks the split symmetry induced by the metric split \eq{eqn:genmetricsplit} in the path integral. Therefore, even though the dependence of the classical action $S[g_{\mu \nu}]$ on $\bg_{\mu \nu}$ and $h_{\mu \nu}$ is directly given by \eq{eqn:genmetricsplit}, the resulting effective average action is a functional whose dependence on $\bg_{\mu \nu}$ and $h_{\mu \nu}$ breaks \eq{eqn:genmetricsplit} and is not known a priori. Thus, we might think of $\Gamma_k$ as a functional with an unrelated dependence on the quantum field $\phi_i$ and the background metric $\bg_{\mu \nu}$,
\begin{equation}
	\Gamma_k = \Gamma_k [\overline{g}_{\mu \nu}, \phi_i] \, .
\end{equation}
Here, $\phi_i$ includes the fluctuation metric $h_{\mu \nu}$ as well as ghost fields such as arising from gauge fixing or the measure of the path integral. In principle, the relation between the dependence on $\bg_{\mu \nu}$ and $h_{\mu \nu}$ can be obtained using modified split symmetries. For alternative strategies, see \cite{Pawlowski:2020qer}.

In our approach we will follow the single field or background field approximation. The basic intuition for this approximation lies in the idea that the flow remains to be driven by objects not modifying the metric split which is induced in \eq{eqn:genmetricsplit}. Doing so, we write the action as
\begin{equation}
	\Gamma_k [\overline{g}_{\mu \nu}, \phi_i] = \overline{\Gamma}_k [\overline{g}_{\mu \nu} + \delta g_{\mu \nu}] + \hat{\Gamma}_k [\overline{g}_{\mu \nu}, \phi_i] \, ,
\end{equation}
with 
\begin{equation}
	\hat{\Gamma}_k[\bg_{\mu \nu}, 0] = 0 \, .
\end{equation}
In this notation, $\overline{\Gamma}_k$ only depends on the full metric $g_{\mu \nu}$ and contains only operators not modifying the split of the metric. The background field approximation then amounts to taking $\hat{\Gamma}_k$ to be given by the terms in the bare action arising from the functional measure. Moreover, their $k$-dependence is neglected. Using \eq{eqn:wetterich} we then determine the flow of $\overline{\Gamma}_k$ by evaluating the right-hand-side at vanishing quantum fields, i.e.
\begin{equation}
	\partial_t \overline{\Gamma}_k [\overline{g}_{\mu \nu}] \left. = \frac{1}{2} \text{Tr} \left\{ \frac{\partial_t \mathcal{R}_k}{\Gamma_k^{(2)} + \mathcal{R}_k} \right\} \right|_{\phi_i = 0} \, .
	\label{eqn:wetterichbga}
\end{equation}
An advantage of the background field approximation is that allows the study of rather general types of gravitational theories.

\subsection{Maximally Symmetric Backgrounds}\label{MSB}
An important ingredient of the flow equation is the operator trace in \eq{eqn:wetterichbga}. In principle, curved or flat background geometries can be used for its evaluation. The use of  {\it general} backgrounds, albeit desirable, is out of reach presently for the types of theories considered here. Progress can be made by using maximally symmetric background geometries, whose  simplifications make the evaluation of operator traces  tractable. Equally important, for maximally symmetric backgrounds, all heat kernel coefficients for scalar, vectors, and tensors are available in closed form \cite{Kluth:2019vkg}. This ensures from the outset that polynomial expansions of  Lagrangians can always be performed.

On a maximally symmetric background, all curvature invariants can be expressed in terms of the scalar Ricci curvature which itself is related to the single dimensionful parameter available on such backgrounds. For positive curvature, the resulting geometry is a sphere and the dimensionful parameter its radius. Irrespective of the sign of the curvature we have the following identities,
\begin{equation}
	\bR_{\rho \sigma \mu \nu} = \frac{\bR}{d (d- 1)} \left( \bg_{\rho \mu} \bg_{\sigma \nu} - \bg_{\rho \nu} \bg_{\sigma \mu} \right)\, , \qquad \bR_{\mu \nu} = \frac{\bg_{\mu \nu}}{d} \bR \, , \qquad \bnabla_\mu \bR = 0 \, ,
	\label{eqn:fsbcurv}
\end{equation}
where we have indicated objects constructed from the background metric by a bar. Due to \eq{eqn:fsbcurv}, the only remaining objects which can carry indices are the background metric, quantum fields in the path integral and covariant derivatives acting on them. In the next subsection we see that this combined with a useful field decomposition leads to the absence of non-minimal differential operators in functional traces, i.e. on the right-hand side of \eq{eqn:wetterichbga}.

\subsection{Field Decomposition}
Following \eq{eqn:wetterichbga}, we require the Hessians of $\Gamma_k$ evaluated on the background geometry to compute the flow. The usage of a maximally symmetric background gives the most general form of such Hessians as
\begin{equation}
	\left. \fderiv{\Gamma_k}{\phi_i \delta \phi_j} \right|_{\phi_\ell = 0} = \sum_n u_n \left(\bR\right) v_n \left(\bg_{\mu \nu}, \bnabla \right) \, ,
\end{equation}
with $u_n (\bR)$ a scalar only depending on the background scalar curvature and $v_n (\bg_{\mu \nu}, \bnabla)$ a tensor carrying the Lorentz indices of the fields $\phi_i$ and $\phi_j$. At this point, $v_n$ might include differential operators with indices of tensor fields and is not a function of the Laplacian only. To ensure that $v_n$ can be written as a function of the Laplacian, we need to decompose all fields carrying indices into transverse and traceless pieces. For a vector field $T_\mu$, this decomposition is well known and given by
\begin{equation}
	T_\mu = \xi_\mu + \bnabla_\mu \eta \, , \qquad \text{with} \qquad \bnabla^\mu \xi_\mu = 0 \, .
	\label{eqn:vectordecomp}
\end{equation}
Note that the scalar field $\eta$ is fully determined up to a constant shift which drops out due to the covariant derivative. Hessians between the fields $\xi_\mu$ and $\eta$ can only include the Laplacian after commuting covariant derivatives. The reason for this is that a covariant derivative in $v_n$ carrying an open index would spoil the transverseness required after the field decomposition \eq{eqn:vectordecomp}. 
From this consideration it also follows that the Hessian between fields containing a different number of indices must vanish.

There is an analogous decomposition for a general symmetric tensor fields $h_{\mu \nu}$ given by the York decomposition \cite{York:1973ia},
\begin{equation}
	h_{\mu \nu} = h^T_{\mu \nu} + \bnabla_\mu \xi_\nu + \bnabla_\nu \xi_\mu + \left( \bnabla_\mu \bnabla_\nu - \frac{\bg_{\mu \nu}}{d} \bnabla^2 \right) \sigma + \frac{\bg_{\mu \nu}}{d} h \, ,
	\label{eqn:yorkdecom}
\end{equation}
with 
\begin{equation}\label{eqn:yorkdecom2}
	h^T_{\mu \nu} = h^T_{\nu \mu}  \, , \qquad \bnabla^\mu h^T_{\mu \nu} = 0 \, , \qquad \bg^{\mu \nu} h^T_{\mu \nu} = 0 \, , \qquad \bnabla^\mu \xi_\mu = 0 \, .
\end{equation}
Note that the different York modes are given unambiguously up to Killing vectors $\xi_\mu$, constant scalars $\sigma$, and conformal Killing vectors $\nabla_\mu \sigma$ whose contributions drop out from \eq{eqn:yorkdecom}. By the same arguments as before, it follows that Hessians between these York modes can only include minimal differential operators. It also follows that Hessians between fields containing a different number of indices vanish.

Using the York decomposition, the Hessian of $\Gamma_k$ w.r.t. $h_{\mu \nu}$ becomes matrix-valued; the components of this matrix refer to the different York modes. Using the simplifications just discussed, its most general form is given by
\begin{equation}
	\Gamma_k^{(2)} = \begin{pmatrix}
		\Gamma_k^{h^T h^T} & 0 & 0 & 0 \\
		0 & \Gamma_k^{\xi \xi} & 0 & 0 \\
		0 & 0 & \Gamma_k^{\sigma \sigma} & \Gamma_k^{\sigma h} \\
		0 & 0 & \Gamma_k^{h \sigma} & \Gamma_k^{h h}
	\end{pmatrix} \, ,
	\label{eqn:Hessianmatrix}
\end{equation}
on a maximally symmetric background.\footnote{Note that here we only consider the part of the Hessian describing contributions associated to the York modes.} This matrix is diagonal apart from the $2 \times 2$ submatrix between $\sigma$ and $h$. 

The York decomposition \eq{eqn:yorkdecom} as well as the decomposition for vector fields \eq{eqn:vectordecomp} are background metric dependent redefinitions of fields integrated over in the path integral. As such, they introduce non-trivial Jacobians into the measure of the path integral. We take care of these using the Faddeev-Popov trick and write them as contributions to $\hat{\Gamma}_k$ arising from ghost fields. For details of this procedure the reader is referred to \cite{Falls:2017lst}.

\subsection{Gauge Fixing}
\label{sct:gauge}
Next, we need to gauge fix the gravitational action to make the propagator invertible. We use a standard gauge fixing action given by
\begin{equation}
	\Gammagf = \frac{1}{2 \alpha} \grdint{d}{x} \mathcal{F}_\mu \mathcal{F}^\mu \, ,
\end{equation}
with
\begin{equation}
	\mathcal{F}_\mu = \sqrt{2} \kappa \left( \bnabla^\nu h_{\mu \nu} - \frac{1 + \delta}{4} \bnabla_\mu \overline{h} \right) \, .
	\label{eqn:gfFgen}
\end{equation}
The parameters $\alpha$ and $\delta$ are arbitrary gauge parameters and are often chosen to help with computational simplicity. 

A  useful gauge choice is given by the Landau gauge $\alpha \rightarrow 0$, as it was argued in \cite{Litim:1998qi} that it is a renormalisation group fixed point for the gauge parameters. This also serves as a justification for neglecting the renormalisation group running of the gauge parameter. While this restricts the parameter $\alpha$, $\delta$ is not fixed by a similar argument. Technically however, there are two useful choices, namely $\delta = 0$ and $\delta \rightarrow \infty$, the latter one also known as the unimodular gauge. In the first case ($\delta = 0$) and on maximally symmetric backgrounds, the Hessian of the gauge fixing action becomes
\begin{equation}
	\begin{split}
		G^{(2)} =& \, \frac{\kappa^2}{16 \alpha} \int \text{d}^4 x \sqrt{\bg} \, \left[ \xi_\mu \left( \bR + 4 \bnabla^2 \right)^2 \xi^\mu - \sigma \bnabla^2 \left( \bR + (d - 1) \bnabla^2 \right)^2 \sigma \right] \, .
	\end{split}
\end{equation}
In the Landau gauge together with $\delta = 0$, contributions from $\bGamma_k$ to the Hessians involving $\xi_\mu$ or $\sigma$ can be neglected, simply because the contributions from $G^{(2)}$ are dominant due to $\alpha \rightarrow 0$. This even remains true for the off-diagonal elements contained in \eq{eqn:Hessianmatrix} when using \eq{eqn:wetterichbga}.  For generic $\delta$, however, such a simplification does not occur, and it is necessary to consider Hessians coming from $\bGamma_k$ together with Hessians from the gauge fixing action. The only other choice for $\delta$ leading to simplifications is given by the unimodular gauge. In this case, the Hessian of $G$ remains unchanged for $G^{\xi \xi}$, however, its dominant contributions to the $2 \times 2$ submatrix in $\eq{eqn:Hessianmatrix}$ are located in the Hessian of $G^{h h}$.  It follows that contributions from $\bGamma_k$ to Hessians of $\xi_\mu$ or $h$ can be neglected in the unimodular gauge.

Even though either of these gauge choices  lead to welcome simplifications of the flow equation, we focus below on the Landau gauge with  $\delta = 0$.
 From a practical point of view, this can, at least  partly, be motivated by invoking a principle of least 
variation, observed in the Einstein-Hilbert theory \cite{Gies:2015tca}, which also favours  the exponential split together with $\delta = 0$.
Also, this choice together with a linear split of the metric $(\tau = 0)$  has been adopted by many previous works in the literature, offering
points of contact for consistency checks of results.

Finally, just as the York decomposition, this gauge fixing procedure introduces a non-trivial determinant into the path integral. This is taken care of using the Faddeev-Popov trick in the same way as the Jacobian arising from the York decomposition \cite{Falls:2017lst}.

\subsection{Flows for Quantum Gravity}

Having specified the gauge fixing in the previous section, we are now able to explicitly invert the matrix in field space in \eq{eqn:wetterichbga}. For this, let us remember that there are four York modes entering the flow ($h^T_{\mu \nu}$, $\xi_\mu$, $\sigma$, and $h$) as well as various ghost fields arising from the Faddeev-Popov procedure and the Jacobians induced by the York decomposition. Without going into details we note that these ghost fields include 5 transverse vectors and 7 scalars.\footnote{We count Graßmann variables and their complex conjugate as separate fields.} Under the assumption that the regulator terms take the same form as the Hessian and in particular making use of \eq{eqn:Hessianmatrix}, \eq{eqn:wetterichbga}, the flow equation  boils down to
\begin{equation}\label{prefRG}
	\begin{split}
		\partial_t \bGamma_k =& \, \frac{1}{2} \text{Tr}_{2} \left\{ \frac{\partial_t \mathcal{R}_k^{h^T h^T}}{\Gamma_{k}^{h^T h^T} + \mathcal{R}_k^{h^T h^T}} \right\} - \frac{1}{2} \text{Tr}_{1}{}^{(}{}'{}^{)} \left\{ \frac{\partial_t \mathcal{R}^{V}_k}{-\nabla^2 - \frac{R}{d} + \mathcal{R}^{V}_k} \right\} \\
		& - \frac{1}{2} \text{Tr}_{0}{}^{(}{}''{}^{)} \left\{ \frac{\partial_t \mathcal{R}^{S}_k}{- \nabla^2 - \frac{R}{d - 1} + \mathcal{R}^{S}_k} \right\} + \frac{1}{2} \text{Tr}_{0} \left\{ \frac{\partial_t \mathcal{R}_k^{h h}}{\Gamma^{h h}_k + \mathcal{R}_k^{h h}} \right\} \, ,
	\end{split}
\end{equation}
in Landau gauge with $\delta = 0$. In accordance with the background field approximation it is implicitly understood that fluctuation fields are set to zero after computing the Hessians.
These traces are functional traces over fields of different spin as indicated by their subscripts. Hence, the traces $\text{Tr}_0$, $\text{Tr}_1$, and $\text{Tr}_2$ are understood as traces over scalars, transverse vectors, and transverse traceless symmetric tensors, respectively. Moreover, primes denote the exclusion of lowest modes which should be excluded if they do not contribute to the field decomposition into transverse and traceless pieces \cite{Lauscher:2001ya}. Note that these modes should only be excluded on maximally symmetric backgrounds with positive curvature. For this reason, we have put the primes in brackets to indicate that an exclusion is not necessary on a hyperbolic space \cite{Falls:2016msz}.

The first and the last trace of \eq{prefRG} are directly related to the fluctuations of the transverse mode $h^T_{\mu \nu}$ and the trace mode $h$. As such, they only receive contributions from the physical part $\bGamma_k$ of the effective average action. The fluctuations w.r.t. $\xi_\mu$ as well as the five different transverse vector ghosts are contained in the second trace. To arrive at this compact result we have regulated all these contributions with the same regulator $\mathcal{R}_k^{V}$. Due to the gauge fixing contributions being dominant over $\bGamma_k^{\xi \xi}$ in Landau gauge it follows that the second trace is completely independent of $\bGamma_k$. Similarly, all contributions from the $\sigma$ mode and the seven remaining  different scalar ghost contributions are contained in the third trace. Again, for the chosen gauge all contributions from $\bGamma_k$ to the Hessian of $\sigma$ can be neglected as well as the off-diagonal elements in \eq{eqn:Hessianmatrix}. Using the same regulator $\mathcal{R}_k^{S}$ for these contributions we arrive at \eq{prefRG}; see  \cite{Falls:2017lst,Falls:2018ylp} for more details.

\subsection{Wilsonian Cutoff}
The next ingredient to the functional renormalisation group is the infrared regulator, which has to be introduced for each field in the path integral. Thus far, we have not made any assumptions about its explicit shape or whether 
it depends on couplings in the effective action,  simply because the form of \eq{prefRG} does not depend on such choices.
In the following we aim at finding simple analytic flows, also guided by stability considerations.
  Still, we emphasise that our regulator choices are by no means mandatory, and perfectly viable and tractable flows
  can be found for other choices.

The first choice we make is defining the regulator \eq{prefRG} by the replacement rule  \cite{Codello:2008vh}
\begin{equation}
	\Gamma_k^{\phi_i \phi_j} \left( -\bnabla^2 \right) + \mathcal{R}^{\phi_i \phi_j}_k \left( - \bnabla^2 + E_i \right) = \Gamma_k^{\phi_i \phi_j} \left( R_k \left( - \bnabla^2 + E_i \right)  -\bnabla^2 \right) \, ,
	\label{eqn:regrep}
\end{equation}
for physical contributions depending on $\bGamma_k$.  The shape function $R_k(q^2)$ obeys the limits $R_k(q^2)>0$ for $q^2/k^2\to 0$ and $R_k(q^2)\to 0$ for $k^2/q^2\to 0$ \cite{Litim:2000ci}.  We use the optimised cut-off \cite{Litim:2001up}
\begin{equation}
	R_k (z) = (k^2 - z) \theta (k^2 - z) \,,
	\label{cutofffunc}
\end{equation}
which leads to simple, analytical flows with  enhanced convergence   properties \cite{Litim:2001fd,Litim:2002cf,Litim:2002qn}, e.g.~in the local potential approximation (LPA) which is similar to the approximations considered here for gravity.
The parameters $E_i$ in \eq{eqn:regrep} are endomorphisms which can be chosen freely, subject to positive definiteness of the resulting Laplacian,
\begin{equation}
	- \bnabla^2 + E_i > 0 \,.
	\label{eqn:laplaceposdef}
\end{equation} 
 In general, we might introduce different endomorphisms for different contributions in \eq{prefRG}. Doing so, we denote the endomorphism in the regulator for the transverse tensor modes by $E_1$ and the endomorphism for the trace mode by $E_4$. For the remaining regulators we choose
\begin{equation}
	\mathcal{R}_k^{V} = R_k \left( - \nabla^2 + E_2 \right) \, , \qquad \mathcal{R}_k^{S} = R_k \left( - \nabla^2 + E_3 \right) \, .
\end{equation}
The resulting bounds on the endomorphism parameters from \eq{eqn:laplaceposdef} can be read off from the eigenvalues of $- \nabla^2$ acting on fields of different spin and requiring that all eigenvalues stay positive after adding the endomorphism. Taking into account that some modes are excluded from the functional traces this yields
\begin{equation}
	\label{eqn:endobounds}
	\begin{aligned}
		E_1 >& \, - \frac{2 (2 + d - 1) - 2}{d (d - 1)} R \, , \qquad& E_2 >& \, - \frac{2 (2 + d - 1) - 1}{d (d - 1)} R \, , \\
		E_3 >& \, - \frac{2 (2 + d - 1)}{d (d - 1)} R \, , \qquad& E_4 >& \, 0 \, ,
	\end{aligned}
\end{equation}
for positive curvature backgrounds.

Note that the optimised cut-off \eq{cutofffunc} vanishes identically whenever $- \bnabla^2 + E_i > k^2$. It follows that the propagators in \eq{prefRG} are only non-zero when $- \bnabla^2 + E_i \leq k^2$. In this regime, the Heaviside function in \eq{cutofffunc} is unity and all propagators are effectively rendered constant. All in all, these simplifications lead to
\begin{equation}\label{fRG}
	\begin{split}
		\partial_t \bGamma_k =& \, \frac{1}{2} \text{Tr}_{2} \left\{ \frac{\partial_t \mathcal{R}_k^{h^T h^T} \left( - \nabla^2 + E_1 \right)}{\Gamma_k^{h^T h^T} ( k^2 - E_1 )} \right\}  - \frac{1}{2} \text{Tr}_{1}' \left\{ \frac{\partial_t R_k \left( -\nabla^2 + E_2 \right)}{k^2 - E_2 - \frac{R}{d}} \right\} \\
		& - \frac{1}{2} \text{Tr}_{0}'' \left\{ \frac{\partial_t R_k \left( - \nabla^2  + E_3 \right)}{k^2 - E_3 - \frac{R}{d - 1}} \right\} + \frac{1}{2} \text{Tr}_{0} \left\{ \frac{\partial_t \mathcal{R}_k^{h h} \left( - \nabla^2 + E_4 \right)}{\Gamma_k^{h h} ( k^2 - E_4 )} \right\} \, ,
	\end{split}
\end{equation}
on spherical backgrounds.
Hence, the only differential operators we need to take care of in the functional traces are polynomials in $-\bnabla^2$ coming from $\partial_t \mathcal{R}^{\phi_i \phi_j}_k$ multiplied by Heaviside functions from the optimised cut-off \eq{cutofffunc}. In particular, no inverse of a differential operator or linear combinations thereof is required.

\subsection{Trace Technology}

Lastly, we need to consider the computation of the functional traces in \eq{fRG}. As noted already, the only differential operators we have to deal with are  Laplacians to non-negative powers multiplied by Heaviside theta functions arising from the optimised cut-off, i.e.~functions of the form
\begin{equation}
	W_n \left(- \bnabla^2\right) \equiv \left(- \bnabla^2\right)^n \theta (k^2 + \bnabla^2 - E) \, .
\end{equation}
We compute these traces using the early time expansion of the heat kernel. This is done by first using the anti-Laplace transformation to write
\begin{equation}
	{\rm Tr}_\text{spin} \left\{ W_n \left( - \bnabla^2 \right) \right\} = \int_0^\infty \text{d} s \, \widetilde{W}_n (s) \text{Tr}_\text{spin} \left\{ e^{s \bnabla^2} \right\}
\end{equation}
The general form of the heat kernel expansion on spheres is given by \cite{Kluth:2019vkg}
\begin{equation}
	{\rm Tr}_\text{spin} \left\{ e^{s \bnabla^2} \right\} = \frac{\text{Vol}}{(4 \pi s)^{d/2}} \sum_{n = 0}^\infty \left[ b_{2n}^{(\text{spin})} s^n + c_{d + 2n}^{(\text{spin})} s^{d/2 + n} \right] \, .
\end{equation}
Therefore,
\begin{equation}
	\begin{split}
		{\rm Tr}_\text{spin} \left\{ W_n \left( - \bnabla^2 \right) \right\} =& \, \frac{\text{Vol}}{(4 \pi)^{d/2}} \int_0^\infty \text{d} s \, \sum_{m = 0}^\infty \widetilde{W}_n (s) \left[ b_{2m}^{(\text{spin})} s^{m - d/2} + c_{d + 2m}^{(\text{spin})} s^{m} \right] \\
		=& \, \frac{\text{Vol}}{(4 \pi)^{d/2}} \int_0^{k^2 - E} \text{d} \omega \, \sum_{m = 0}^\infty \left[ \frac{b_{2m}^{(\text{spin})}}{\Gamma (d/2-m)} \omega^{n + d/2 - 1 - m} + \frac{c_{d + 2m}^{(\text{spin})}}{\Gamma(-m)} \omega^{n - 1 - m} \right] \\
		=& \, \frac{\text{Vol}}{(4 \pi)^{d/2}} \sum_{m = 0}^\infty \left[ \frac{b_{2m}^{(\text{spin})} k^{2 (n + d/2 - m)}}{(n + d/2 - m) \Gamma(d/2 - m)} + \frac{c_{d + 2m}^{(\text{spin})} k^{2 (n - m)}}{(n - m) \Gamma(- m)} \right] \, ,
	\end{split}
	\label{HKres}
\end{equation}
where we have used
\begin{equation}
	s^n = \frac{1}{\Gamma(-n)} \int_0^\infty \text{d} \omega \, \omega^{-1 -n} e^{-\omega s} \, .
\end{equation}
Note that in obtaining \eq{HKres} we have used analytical continuation in $d$. The coefficients $c^{(2)}_{d + 2m}$ are only non-vanishing for fields fulfilling differential constraints, i.e.~in the cases $\text{spin} = 1$ and $\text{spin} = 2$ and originate from excluded modes. 

For large $m$ and even dimension, the $\Gamma$-functions in \eq{HKres} can become singular. Due to the presence of these poles, we note that in even dimensions only a finite number of heat kernel coefficients is required. This is due to the fact that for large enough $m$, the poles induced by the Gamma functions in the denominators cannot be compensated by anything else in these equations wherefore these contributions vanish. The last non-vanishing contributions are given by 
\begin{equation}
	n + \frac{d}{2} - m = 0 \, , \qquad n - m = 0 \, ,
\end{equation}
for the $b$ and $c$ coefficients, respectively. In these cases, the poles are compensated by zeros in the denominators of \eq{HKres} and give a finite contribution.
The fact that only a finite number of terms contribute in \eq{HKres} can be traced back to the properties of the optimised cut-off and leads to only a finite number of heat kernel coefficients contributing to the flow equation. Even though this choice leads to a somewhat simpler structure, it is not required to stick to it since all heat kernel coefficients on spheres are known \cite{Kluth:2019vkg}. For this reason,  it is possible to obtain explicit flows for  generic cut-off functions other than \eq{cutofffunc}.

This concludes our algorithm to compute flow equations of the form \eq{eqn:wetterichbga}. What is left is the form of the Hessians of $\bGamma_k$ contributing to the flow. This is the subject of the next section.

\section{\bf Higher Order Theories of Gravity}
\label{sct:Bueno}
In this section, we derive functional renormalisation group flows for $f(R_{\mu\nu\rho\sigma})$ type  theories of gravity, whose actions 
are general functions of the Riemann tensor and the inverse metric. 
We  also explain the role of maximally symmetric backgrounds, which are used for  the determination of operator 
traces.

\subsection{Action}
From now on we assume that the Lagrangian $\mathcal{L} $ is a general function of the Riemann tensor and the inverse metric, without  any  covariant derivatives acting on Riemann tensors. The gravitational actions  can therefore be written as
\begin{equation}
	\bGamma_k [g_{\mu \nu}] = \int \text{d}^d x \sqrt{g} \ \mathcal{L}  ( R_{\rho \sigma \mu \nu}, g^{\alpha \beta} )\,. 
	\label{eqn:bGammaform} 
\end{equation}
 This covers a rather wide range of models including the Einstein-Hilbert theory, Stelle's fourth order theory for gravity, $f(R)$ models, and many more higher order extensions of gravity. We also note that the search for asymptotically safe fixed points of quantum gravity has  almost exclusively been focussed on specific models of the type  \eq{eqn:bGammaform}.

Flow equations for actions \eq{eqn:bGammaform} are particularly useful when considering LPA-like approximations for gravity in the spirit of \cite{Benedetti:2012dx}, see  e.g.~\cite{ Reuter:1996cp,Lauscher:2002sq,Litim:2003vp,Machado:2007ea,Codello:2008vh,  Benedetti:2009rx,Falls:2013bv,
Falls:2014tra, Falls:2018ylp,Kluth:2020bdv,Falls:2020qhj}.
Selecting one  curvature invariant  per mass dimension which is   non-vanishing on maximally  symmetric backgrounds, we may expand the action into a power series of curvature invariants,
\begin{equation}
	\bGamma_k = \sum_{n = 0}^\infty \int \text{d}^d x \sqrt{g} \ \blambda_n\, X_n  ( R_{\rho \sigma \mu \nu}, g^{\alpha \beta} )\,.
	\label{eqn:ansatzgenfsb}
\end{equation}
Here $n\ge 0$ sums over operators $X_n$ with canonical mass dimension $[X_n]=2n$,  constructed out of the Riemann tensor and the inverse metric, and associated coupling constants $\blambda_n$ with canonical mass dimension $[\blambda_n]=d-2n$.

In the remainder, we  derive general flow equations for actions \eq{eqn:bGammaform} or polynomial couplings as in \eq{eqn:ansatzgenfsb}. From a practical point, we take the view that $\{R_{\rho \sigma \mu \nu}, g^{\alpha \beta} \}$ are the fundamental variables of $\mathcal{L}$. 
Alternative choices for the fundamental variables such as $\{R^{\rho \sigma}_{ \mu \nu} \}$ or $\{R^{\rho}_{ \ \ \sigma \mu \nu} , g^{\alpha \beta}\}$ can be taken as well and would, at best, change intermediate algebraic expressions without affecting the final outcome \cite{Padmanabhan:2011ex}.

\subsection{First and Second Variations}
In order to study quantum effects for actions of the type  \eq{eqn:bGammaform} with the help of  functional renormalisation \eq{eqn:wetterichbga}, \eq{fRG},  we must provide the second variation of the action. In general, it is given by
\begin{equation}
	\delta^2  \,\bGamma_k = \int {\rm d}^d x \left[ \mathcal{L} \,\delta^2 \sqrt{g}+ 2\, \delta \sqrt{g} \, \delta \mathcal{L}  + \sqrt{g} \, \delta^2 \mathcal{L}  \right] \,.
	\label{eqn:bGammavar2}
\end{equation}
Here and in the following it is  understood that the metric is split  into a background and a fluctuation field according to \eq{eqn:genmetricsplit}, and the fluctuation field $h_{\mu \nu}$ is set to zero after computing the variations. Next, we account for the fact that $\mathcal{L}$ is taken to be a function of the Riemann tensor and the inverse metric. 
Introducing
\begin{equation}\label{W}
	\mathcal{W}^{\rho \sigma \mu \nu} \equiv \pderiv{\mathcal{L}}{R_{\rho \sigma \mu \nu}} \,,
\end{equation}
to denote the Riemann tensor derivative of the Lagrangian, we write its first variation  as
\begin{equation}\label{First}
		\delta \mathcal{L}  = \, \mathcal{W}^{\rho \sigma \mu \nu} \,\delta R_{\rho \sigma \mu \nu} 
		+ \pderiv{\mathcal{L}}{g^{\mu \nu}}\, \delta g^{\mu \nu} \,.
		\end{equation}
Similarly, the second variation reads	
\begin{equation}
\begin{split}
		\delta^2 \mathcal{L} =& \, \mathcal{W}^{\rho \sigma \mu \nu}\, \delta^2 R_{\rho \sigma \mu \nu}
		+ \pderiv{\mathcal{W}^{\rho \sigma \mu \nu}}{R_{\alpha \beta \gamma \delta}}\, \delta R_{\rho \sigma \mu \nu}\, \delta R_{\alpha \beta \gamma \delta} 
		+ 2 \pderiv{\mathcal{W}^{\rho \sigma \mu \nu}}{g^{\alpha \beta}}\, \delta R_{\rho \sigma \mu \nu}\, \delta g^{\alpha \beta} \\
		& + \pderiv{\mathcal{L}}{g^{\mu \nu}} \,\delta^2 g^{\mu \nu} 
		+ \pderiv{\mathcal{L} }{g^{\rho \sigma} g^{\mu \nu}}\, \delta g^{\rho \sigma}\, \delta g^{\mu \nu} \,.
	\end{split}
	\label{eqn:firstsecvareq}
\end{equation}
Evidently, both \eq{First} and \eq{eqn:firstsecvareq} involve first and second derivatives with respect to the Riemann tensor and the inverse metric. 
However, it so turns out that all terms involving first or second derivatives with respect to the metric can be re-expressed  in terms of 
\eq{W} and its Riemann tensor derivative. Specifically, the first metric derivative  is found to be proportional to $\mathcal{W}$,
\begin{equation}
	\pderiv{\mathcal{L}}{g^{\lambda \eta}} = 2 g_{\rho ( \lambda} R_{\eta) \sigma \mu \nu} \mathcal{W}^{\rho \sigma \mu \nu} \,.
	\label{eqn:LgdR}
\end{equation}
whereas the second derivatives
\begin{equation}
		\pderiv{^2\mathcal{L}}{R_{\rho \sigma \mu \nu}\partial  g^{\alpha \beta}} 
		=\, g_{(\beta}^{\ \ [ \rho} \mathcal{W}_{\alpha)}^{\ \ \sigma ] \mu \nu} + g_{(\beta}^{\ \ [ \mu} \mathcal{W}_{\alpha)}^{\ \ \nu ] \rho \sigma} + 2 g_{\zeta (\alpha} R_{\beta) \kappa \eta \xi} \pderiv{\mathcal{W}^{\zeta \kappa \eta \xi}}{R_{\rho \sigma \mu \nu}} \,,
	\label{eqn:LgdRdR}
\end{equation}
\begin{equation}
	\pderiv{^2 \mathcal{L}}{g^{\rho \sigma} \partial g^{\mu \nu}} = - 2 g_{\alpha (\mu} g_{\nu) (\rho} \mathcal{W}^{\alpha \beta \gamma \delta} R_{\sigma) \beta \gamma \delta} + 2\, 
	\pderiv{^2\mathcal{L}}{R_{\alpha \beta \gamma \delta}\partial  g^{\mu \nu}} 
	\,g_{\alpha (\rho} \,R_{\sigma) \beta \gamma \delta} 
	\label{eqn:LgdRdRdg}
\end{equation}
are linear in $\mathcal{W}$ and its first Riemann derivative. The detailed  derivation of  \eq{eqn:LgdR}, \eq{eqn:LgdRdR} and \eq{eqn:LgdRdRdg}   is  relegated to  App.~\ref{sct:metricderiv}. We therefore conclude that the first and second variations require the knowledge of $\mathcal{W}$ and its partial derivative  $\partial \mathcal{W}/\partial R_{ \mu \nu\rho \sigma}$ for general background.

\subsection{Maximally Symmetric Backgrounds}
\label{sct:riemderiv}
In this section we explain why $\mathcal{W}$ and its partial derivative  $\partial \mathcal{W}/\partial R_{ \mu \nu\rho \sigma}$ are uniquely determined 
in terms of  a few scalar functions without specifying the underlying Lagrangian, provided maximally symmetric backgrounds are used  \cite{Bueno:2016xff}.

The basic observation is that tensors on maximally symmetric backgrounds, characterised by \eq{eqn:fsbcurv}, can only be constructed from the metric tensor and  functions of the background  scalar curvature $R$. Furthermore, derivatives with respect to  Riemann tensors inherit the symmetries of the Riemann tensor.  
With these requirements in mind, we observe that the action, evaluated on a maximally symmetric background, is characterised by a scalar function of the  Ricci scalar curvature,
\begin{equation}\label{Ldef}
	 \mathcal{L} ( R_{\rho \sigma \mu \nu}, g^{\alpha \beta} ) \Big|_{\rm msb} =L\,,
\end{equation}
where it is understood that $L=L(R)$.  The specific  form of  $L$ is unknown presently as it evidently depends on the choice for the action $\mathcal{L}$.
 
 By the same token, the first Riemann derivative of the Lagrangian on maximally symmetric backgrounds takes the form
\begin{equation}
	 \left. \pderiv{\mathcal{L}}{R_{\rho \sigma \mu \nu}}\right|_\text{msb}  \equiv  \left.\mathcal{W}^{\rho \sigma \mu \nu} \right|_\text{msb} = E \,\mathcal{P}^{\rho \sigma \mu \nu}
	\label{eqn:firstderivL}
\end{equation}
where  $E=E(R)$ a scalar function of the Ricci scalar curvature.\footnote{Here and in the following we omit the argument of $E$. Its dependence on the Ricci scalar curvature is understood implicitly.} Again, the specific  form of the function $E$ is unknown presently, but would be specified uniquely as soon as  the explicit form of the action $\mathcal{L}$ is provided.
 The tensor
\begin{equation}
\mathcal{P}^{\rho \sigma \mu \nu} = g^{\rho [\mu} g^{\nu] \sigma} 
	\label{Pdef}
\end{equation}
has the symmetries of the Riemann tensor, and can also be understood as $\partial R/\partial R_{\rho \sigma \mu \nu}$.

Finally, the second Riemann tensor derivative of the Lagrangian must contain tensor structures constructed from the metric fulfilling all symmetries inherited from the Riemann tensor, up to unknown scalar functions of the Ricci scalar curvature. This can be written as
\begin{equation}\label{basis}
	\left. \pderiv{^2 \mathcal{L}}{R_{\rho \sigma \mu \nu} \partial R_{\alpha \beta \gamma \delta}} \right|_{\rm msb} = \sum_n T_n^{\rho \sigma \mu \nu \alpha \beta \gamma \delta} f_n (R) \, ,
\end{equation}
with  tensors $T_n$ constructed from the metric, and $n$ summing over the independent tensors. Based on the properties of the Riemann tensor, the tensors $T_n$ are symmetric in
\begin{subequations}
	\begin{align}
		 \{ \rho, \sigma, \mu,\nu \} &\leftrightarrow \{ \alpha, \beta, \gamma, \delta \} \, ,
		& \{ \rho, \sigma \} &\leftrightarrow \{ \mu, \nu \} \,,
		& \{ \alpha, \beta \} &\leftrightarrow \{ \gamma, \delta \} \,,
	\end{align}
and antisymmetric in
	\begin{align}
		 \rho \leftrightarrow \sigma \,,&
		&\mu \leftrightarrow \nu \,,&
		& \alpha \leftrightarrow \beta \,,&
		&\gamma \leftrightarrow \delta \,,
	\end{align}
	\label{eqn:Riem2symmprop}
\end{subequations}
and should fulfil the algebraic Bianchi identity. There are exactly $n=3$ different non-vanishing tensor structures fulfilling all of these symmetry properties. We write them as
\begin{align}
	\label{Adef}
		\mathcal{A}^{\rho \sigma \mu \nu \alpha \beta \gamma \delta}& = \, \mathcal{P}^{\rho \sigma \mu \nu} 
		\,\mathcal{P}^{\alpha \beta \gamma \delta} \, , \\
\mathcal{B}^{\rho \sigma \mu \nu \alpha \beta \gamma \delta} &= \, \frac{1}{4} \bigg[ g^{\beta] [\rho} g^{\sigma] [\mu} g^{\nu] \gamma} g^{\delta [\alpha} + g^{\sigma] [\alpha} g^{\beta] [\gamma} g^{\delta] \mu} g^{\nu [\rho} 
\label{Bdef}
+ g^{\beta] [\mu} g^{\nu] [\rho} g^{\sigma] \gamma} g^{\delta [\alpha} + g^{\nu] [\alpha} g^{\beta] [\gamma} g^{\delta] \rho} g^{\sigma [\mu} \bigg] \\
	\begin{split}
	\label{Cdef}
		\mathcal{C}^{\rho \sigma \mu \nu \alpha \beta \gamma \delta} &= \, \frac{1}{6} \bigg[ 2 g^{\alpha [\rho} g^{\sigma] \beta} g^{\gamma [\mu} g^{\nu] \delta} + 2 g^{\alpha [\mu} g^{\nu] \beta} g^{\gamma [\rho} g^{\sigma] \delta} 
		- g^{\alpha [\rho} g^{\mu] \beta} g^{\gamma [\nu} g^{\sigma] \delta} - g^{\alpha [\nu} g^{\sigma] \beta} g^{\gamma [\rho} g^{\mu] \delta} \\
		& \qquad - g^{\alpha [\rho} g^{\nu] \beta} g^{\gamma [\sigma} g^{\nu] \delta} - 
		g^{\alpha [\sigma} g^{\mu] \beta} g^{\gamma [\rho} g^{\nu] \delta}\bigg] \,.
	\end{split}
\end{align}
Notice that the tensor  $ \mathcal{A}$ can be viewed as the square of $\partial R/\partial R_{\rho \sigma \mu \nu}$, while  
the tensor $ \mathcal{C}$  is equivalent to $\partial{R^{\rho \sigma \mu \nu}}/\partial{R_{\alpha \beta \gamma \delta}}$.  

To check that no further independent tensor structures exist besides $ \mathcal{A}$, $ \mathcal{B}$ and $ \mathcal{C}$, we observe that there are in total $60$ different tensors  containing eight indices which can be constructed from the metric in such a way that they remain non-vanishing under the anti-symmetrisation, as required by \eq{eqn:Riem2symmprop}. The tensors given in \eq{Adef} -- \eq{Cdef}  contain all $60$ of these structures, thus indicating   that the basis is complete.\footnote{The tensor $\mathcal{C}$ differs from the corresponding one used in \cite{Bueno:2016xff,Bueno:2016ypa} by a further symmetrisation. This ensures that the algebraic Bianchi identity is satisfied.}

We conclude that the most general form for the second Riemann tensor derivative of $\mathcal{L}$ on a maximally symmetric background is given by a linear combination 
of the three tensors  \eq{Adef}, \eq{Bdef}, or \eq{Cdef}, and  we can write \eq{basis} as
\begin{equation}\label{ABCdef}
	\left. \pderiv{^2 \mathcal{L}}{R_{\rho \sigma \mu \nu} \partial R_{\alpha \beta \gamma \delta}} \right|_{\rm msb} 
	= A (R) \, \mathcal{A}^{\rho \sigma \mu \nu \alpha \beta \gamma \delta} 
	+ B (R)\, \mathcal{B}^{\rho \sigma \mu \nu \alpha \beta \gamma \delta} 
	+ C (R)\, \mathcal{C}^{\rho \sigma \mu \nu \alpha \beta \gamma \delta} \,
\end{equation}
with background-curvature dependent coefficients $A, B$ and $C$.\footnote{The parameters $A$, $B$, $C$, $E$ are related to the parameters $a$, $b$, $c$, $e$  in  \cite{Bueno:2016xff,Bueno:2016ypa} as 
	$( A, B, C, E ) = (4 b, 4c, 2 a, 2 e )$.} 
	On the whole, we  are left with five undetermined  functions of the Ricci scalar curvature given by $L$ \eq{Ldef}, $E$ \eq{eqn:firstderivL}, and $A, B$ and $C$ \eq{ABCdef}, which together uniquely characterise any Lagrangian of the form $\mathcal{L} =\mathcal{L} \left( R_{\rho \sigma \mu \nu}, g^{\alpha \beta} \right)$ and its first and second Riemann derivatives on maximally symmetric backgrounds. 
Interestingly, only three of these five functions are independent of each other. To see this, we  use the chain rule to find
\begin{equation}
	\pderiv{L}{R}
	\equiv \pderiv{\mathcal{L}}{R_{\rho \sigma \mu \nu}} \pderiv{R_{\rho \sigma \mu \nu}}{R} \Big|_\text{msb} 
	= E \, ,
	\label{eqn:firstderivrel}
\end{equation}
where we used \eq{eqn:firstderivL} together with
	$\mathcal{P}_{\rho \sigma \mu \nu} \mathcal{P}^{\rho \sigma \mu \nu} = {d (d - 1)}/2$, also noting that the partial derivative $\partial R_{\rho \sigma \mu \nu}/\partial R$ is defined via the Ricci decomposition of the Riemann tensor. 
We conclude that the functions $L$ and $E$ are not independent in that we can always replace $E$ by $L'$. Since $L$ already appears in the second variation, 
this effectively removes one unknown parameter.
Another identity arises from  the second derivative where the chain rule implies
\begin{equation}
	\begin{split}
		\pderiv{^2 L}{R^2} 
		\equiv& \pderiv{^2 \mathcal{L}}{R_{\rho \sigma \mu \nu} \partial R_{\alpha \beta \gamma \delta}} \pderiv{R_{\rho \sigma \mu \nu}}{R} \pderiv{R_{\alpha \beta \gamma \delta}}{R} \Bigg|_{\rm msb} \\
		=& \left( A \cdot \mathcal{A}^{\rho \sigma \mu \nu \alpha \beta \gamma \delta} + B \cdot \mathcal{B}^{\rho \sigma \mu \nu \alpha \beta \gamma \delta} + C \cdot \mathcal{C}^{\rho \sigma \mu \nu \alpha \beta \gamma \delta} \right) \frac{4\mathcal{P}_{\rho \sigma \mu \nu} \mathcal{P}_{\alpha \beta \gamma \delta}}{d^2 (d - 1)^2} \,.
	\end{split}
\end{equation}
Using \eq{Pdef}, \eq{Adef}, \eq{Bdef}, and \eq{Cdef}, and contracting all indices,
\begin{equation}
\begin{split}	\mathcal{A} \cdot \mathcal{P} \mathcal{P} &= {d^2 (d - 1)^2}/{4} \, , \qquad \\
\mathcal{B} \cdot \mathcal{P} \mathcal{P} &= {d (d - 1)^2}/{4} \, , \qquad \\
\mathcal{C} \cdot \mathcal{P} \mathcal{P} &= {d (d - 1)}/{2} \, ,
\end{split}
\end{equation}
 we find
\begin{equation}
	\pderiv{^2 L}{R^2} = A + \frac{2}{d (d - 1)} B + \frac{1}{d} C \, .
	\label{eqn:secderivrel}
\end{equation}
Hence, the four functions $L'', A, B$ and $C$ are linearly dependent, and we can eliminate, say,  $A$ in favour of
 $L$, $B$, and $C$. We conclude that out of the five functions $L, E, A, B$ and $C$, only three  are required to characterise the Lagrangian and its first and second Riemann tensor derivatives unambiguously on maximally symmetric backgrounds. In particular, this provides us with general closed expressions for the Hessians without specifying the Lagrangian. Below, we pick the three functions 
\beq
 \label{LBC}
 L(R)\,, \quad B(R)\,, \quad C(R)
\eeq
 as independent functions to characterise the action and its second variations on maximally symmetric backgrounds.

\subsection{Equations of Motion}
\label{sct:EOM}
Using the results from the previous  sections, we can now provide the equations of motion, which take the form
\begin{equation}
		{\cal E}_{\mu\nu}\equiv\left. \delta{\bGamma_k}/ \delta{g_{\mu \nu}} \right|_{\rm msb} = \s012 T_{\mu\nu}
\end{equation}
where we have also written down the energy momentum tensor due to matter fields,
\begin{equation}
		T_{\mu\nu}=-\frac{2}{\sqrt{g}}
\frac{\delta(\sqrt{g} {\cal L}_{\rm matter})}{\delta g^{\mu\nu}}\,.
\end{equation}
The left-hand side of the equation of motion is determined by the function $L$,
\begin{equation}
	\begin{split}
		{\cal E}_{\mu\nu}=& \, \frac{1}{2} \sqrt{g} g^{\mu \nu} L + \sqrt{g} \mathcal{P}^{\alpha \beta \gamma \delta} \fderiv{R_{\alpha \beta \gamma \delta}}{g_{\mu \nu}} L' - 2 \sqrt{g} \frac{R}{d} g^{\mu \nu} L' 
		=\, \frac{1}{d} \sqrt{g} g^{\mu \nu} \left( \frac{d}{2} L - R L' \right) \,. 
			\end{split}
\end{equation}
In particular, in the absence of matter the equations of motion take the form
\begin{equation}
	2R \pderiv{L}{R} =d\, L \,.
	\label{eqn:genEOM}
\end{equation}
It dictates non-trivial relations amongst the various couplings characterising any given higher order theory of gravity. 
Interestingly, the relation \eq{eqn:genEOM} has a simple interpretation in terms of scaling dimensions. We recall that the canonical mass dimension of the Ricci scalar is $[R]=2$ in any dimension. Then, \eq{eqn:genEOM} states that the scaling dimension of $L$, determined by $2R\partial_R$, {\it exactly} matches its canonical mass dimension $[L]=d$ if, and only if, the equation of motion is satisfied. We rush to add that \eq{eqn:genEOM} should not be viewed as  a differential equation for $L$. Rather, for any given Lagrangian, the isolated solutions $R=R_{\rm dS}$ of \eq{eqn:genEOM} determine the availability of de Sitter or anti de Sitter solutions after analytical continuation to Minkowski signature. 

Despite the rather general form of the Lagrangian,  the equations of motion take a very simple form on maximally symmetric spacetimes. Further, taking into account perturbations on a maximally symmetric background, it is possible to determine the particle content  for general Lagrangians $\mathcal{L}$ \cite{Sisman:2011gz,Senturk:2012yi,Tekin:2016vli,Bueno:2016xff,Bueno:2016ypa}. Some of this can already be read off from the Hessians, to which we turn next.

\subsection{Hessians of Higher Order Gravity}
\label{sct:hessiansb}
We are now in a position to provide the Hessians \eq{eqn:bGammavar2} for higher order theories of gravity in explicit  terms. 
We exploit the findings  for
the first and  second variations of the Lagrangian on maximally symmetric backgrounds  of the previous sections, 
and take  $L$, $B$ and $C$ as the unspecified 
scalar functions \eq{LBC}. Moreover, we employ
the $\tau$-dependent metric split \eq{eqn:taumetricsplit},  and 
the York decomposition \eq{eqn:yorkdecom}, \eq{eqn:yorkdecom2} for the fluctuation field.
The result reads
\begin{equation}
	\begin{split}
		\delta^{2} \bGamma_k  =& \int \text{d}^d x \sqrt{\bg} \, \Bigg\{ h^T_{\mu \nu} \Bigg[ \frac{R^2}{(d - 1) d^2} \left( \frac{B}{d - 1} + 2 C \right) - \frac{R}{d (d - 1)} L' + (\tau -1) \left( \frac{L}{2} - \frac{R}{d} L' \right) \\
		& \qquad \qquad + \left( \frac{(d + 1) R}{d (1 - d)} \left( \frac{B}{d + 1} + C \right) +\frac{L'}{2} \right) \bnabla^2 + \left( \frac{B}{4}+C \right) \bnabla^4 \Bigg] h^T_{\mu \nu}  \\
		& \qquad - 2 \xi_\mu  (\tau -1) \left( \frac{L}{2} - \frac{R}{d} L' \right) \Bigg[ \frac{R}{d} + \bnabla^2 \Bigg] \xi^\mu \\
		& \qquad + \sigma \Bigg[ (\tau -1) \left( \frac{L}{2} - \frac{R}{d} L' \right) \left( \frac{R}{d} - \frac{(1 - d)}{d} \bnabla^2 \right) + \Xi \, \bnabla^2 \Bigg] \bnabla^2 \sigma \\
		& \qquad - 2 h \, \Xi \, \bnabla^2 \sigma\\
		& \qquad + h \, \left[ \Xi  + \frac{(\tau -1+\s0d2)}{d} \left(\frac{L}{2} - \frac{R}{d} L'\right) \right] \, h \Bigg\} \, ,
	\end{split}
	\label{eqn:Hessians}
\end{equation}
where the auxiliary function $\Xi$ is given by
\begin{equation}
	\begin{split}
		\Xi =& \, \frac{R^2}{d^2} \left( \frac{d-3}{(d-1) d} (B - C)  + L'' \right) - \frac{(d-2) R}{2 d^2} L' \\
		&  + \left[ \frac{R}{d^2} \left( \frac{d^2+4 d-20}{4 d} B -\frac{d-4}{d} C + 2 (d-1) L'' \right) -\frac{(d - 2) (d - 1)}{2 d^2} L' \right] \bnabla^2 \\
		&  + \frac{d - 1}{d^2} \left( \frac{d^2-8}{4 d} B + \frac{1}{d} C + (d-1) L'' \right) \bnabla^4\,.
	\end{split}
	\label{eqn:sigmahuniv}
\end{equation}
We report the expressions for the Hessian without using the York decomposition  in App.~\ref{app:Hessian}, for completeness. 
Several comments are in order. 
\begin{itemize}
\item[$(i)$]  {\it Equations of motion.}\\ 
 A number of  terms in the Hessians are proportional to the equations of motion \eq{eqn:genEOM}. We have written \eq{eqn:Hessians} such that $L$ only appears in these terms. Essentially all of them  drop out automatically for the exponential split \eq{eqn:metricexp}, 
the sole exception being the trace-modes $h$. 

\item[$(ii)$]      {\it Hessians in the scalar sector.} \\  
The contributions in the $\sigma \sigma$, $\sigma h$, and $h h$ sectors are very similar and differ only by terms proportional to the equations of motion, with the remaining universal piece $\Xi$ as given in \eq{eqn:sigmahuniv}. This is, however, not equivalent to using the exponential split instead, due to a remaining  extra term  in the  $h h$ sector.

\item[$(iii)$]  {\it Decoupling of auxiliary fields.}\\  
Further, we  observe  that the $\xi_\mu \xi_\nu$ sector is proportional to the equations of motion and that it  vanishes identically for the exponential split.
 For an $f(R)$ model of gravity, this  has previously  been noted in \cite{Ohta:2015fcu}. Our result establishes  that this is 
 valid  much more generally, and   independently of the form of the underlying action. 

\item[$(iv)$] {\it Massive spin-2 degrees of freedom.} \\
We can also infer information about the propagating degrees of freedom directly from the Hessians \eq{eqn:Hessians}. Besides the usual massless spin-2 mode of Einsteinian gravity, higher order extensions of general relativity generically feature   a ghost-like massive spin-2 degree of freedom and an additional scalar \cite{Bueno:2016ypa}. In \eq{eqn:Hessians}, a ghost-like massive degree of freedom makes its appearance due to the $\bnabla^4$ contribution in the transverse traceless modes. However, they will be absent provided that
\begin{equation}
	B + 4 C = 0 \,,
	\label{eqn:ghostdof}
\end{equation}
as can be seen from  \eq{eqn:Hessians}.
Trivial examples for  this are $f(R)$ gravities where $B=0=C$, see Sec.~\ref{fRsec}. For non-trivial examples see Sec.~\ref{sec:paramquad} below.

\item[$(v)$] {\it Massive spin-0 degrees of freedom.} \\
The propagating scalar is related to the $\bnabla^4$ term in the auxiliary function $\Xi$, see \eq{eqn:sigmahuniv}. 
As can be seen from the explicit expression, the scalar does not appear in the spectrum provided that
\begin{equation}
	(d^2-8)B + 4 C + 4d(d-1) L'' =0\,.
	\label{eqn:scadof}
\end{equation}
Note that the conditions \eq{eqn:ghostdof} and \eq{eqn:scadof} are independent of each other in any dimension. Hence, demanding the manifest absence of ghosts 
and the absence of the additional scalar  impose additional constraints,  each reducing the number of independent functions by one.
Einstein-like gravities with only a massless spin-2 degree of freedom are obtained if both \eq{eqn:ghostdof} and \eq{eqn:scadof} are satisfied.

\item[$(vi)$] {\it Cosmological constant.}\\ 
We now turn to the role of the cosmological constant, which, by definition, is encoded in the curvature-independent 
part of $\mathcal{L}$.  
Consequently, it can only contribute to the Hessians via the function $L$, but not  via $B$, nor $C$, nor via  derivatives of $L$. 
If the exponential split \eq{eqn:metricexp} is used, $L$ drops out  from the Hessian \eq{eqn:Hessians} and only its derivatives contribute, with the sole exception  of the trace-mode sector $h h$.
It then  follows that the cosmological constant can only make an appearance on
the right-hand side of the flow equation \eq{eqn:wetterichbga} through the trace-mode fluctuations $h$. 
\item[$(vii)$] {\it Decoupling, and links with unimodular gravity.}\\  
For particular gauge choices (such as the unimodular gauge discussed in  \sct{sct:gauge}), the $h h$ contributions from $\bGamma_k$ are suppressed compared to those arising from the  gauge fixing. In this case the use of  \eq{eqn:metricexp} ensures that the cosmological constant drops out entirely from the right-hand side of the flow. 
This implies that the cosmological constant decouples and no longer influences the running of any other gravitational coupling, akin to unimodular versions of gravity where the cosmological constant becomes non-dynamical and only appears as an integration constant \cite{vanderBij:1981ym}. 
Moreover, its own running will be informed entirely by other couplings. Also, provided they achieve an interacting fixed point under the renormalisation group, it  follows that the scaling dimension associated to  the cosmological constant term is invariably set to 
\beq\label{CCscaling}
\vartheta=-d\,.
\eeq 
The scaling dimension agrees exactly with minus the canonical mass dimension of the cosmological constant term in $d$-dimensional spacetime.
The feature \eq{CCscaling} has been observed already in some $f(R)$ works using these choices. 
Here, the result \eq{CCscaling} is established for  general higher order theories of gravity of the form \eq{eqn:bGammaform}.

\item[$(vii)$] {\it General backgrounds.}\\ 
We close with a remark on the decoupling of the cosmological constant for general backgrounds
beyond the maximally symmetric ones used here.
Using once more the interpolating metric split \eq{eqn:taumetricsplit}, and taking the second variation of 
the cosmological constant term $\sim \lambda_0$, we find 
\begin{equation}
	\delta^2 \left( \int \text{d}^d \sqrt{g} \lambda_0 \right) = \int \text{d}^d x \sqrt{g} \left[ \frac{1}{4} h h - \frac{1}{2} h_{\mu \nu} h^{\mu \nu} + \frac{\tau}{2} h_{\mu \nu} h^{\mu \nu} \right] \lambda_0 \, .
\end{equation}
It states that for any $\tau \neq 1$ a non-vanishing cosmological constant   triggers fluctuations in both the trace and in the tensor modes,
thereby leaving a trace in the beta functions for all gravitational couplings, and irrespective of the chosen background geometry.
For $\tau = 1$, however, fluctuations are only generated in the trace mode, again irrespective of the chosen background geometry. 
This can  be seen as a   hint for the irrelevancy of the cosmological constant, and for a potential equivalence between unimodular gravity, 
and the unimodular gauge of standard gravity in non-linear splits of the metric field $(\tau=1)$, as conjectured in  \cite{deBrito:2020rwu}.
\end{itemize}

\subsection{Mapping Actions to Characteristic Functions}
\label{sct:calcparam}
Thus far, it has been established  that the Hessians of a general higher order theory of gravity with action \eq{eqn:bGammaform} are fully determined by three  scalar functions, say $L $, $B$,  and $C$. The latter  depend  on the form of the Lagrangian $\mathcal{L}$ and need to be determined separately for any given action. Here, we present a highly efficient  algorithm to determine the functions $L$, $A$, $B$, $C$, or $E$.

Starting with a Lagrangian $\mathcal{L}(R_{\rho \sigma \mu \nu}, g^{\alpha \beta})$, the main idea  of the algorithm put forward by Bueno and Cano \cite{Bueno:2016xff} consists of introducing a modified Riemann tensor characterised by a free parameter $\alpha$. Concretely, it is given by
\begin{equation}
	\widetilde{R}_{\rho \sigma \mu \nu} = R_{\rho \sigma \mu \nu} + \frac{2 \alpha}{d (d - 1)} \chi_{\rho [\mu} \chi_{\nu] \sigma} \,,
	\label{eqn:Riemannchi}
\end{equation}
with a tensor $\chi_{\mu \nu}$  fulfilling
\begin{equation}
	\chi^\mu_\mu = \chi \, , \qquad \chi_{\mu \alpha} \chi^\alpha_{\,\,\, \nu} = \chi_{\mu \nu} \, , \qquad \chi_{\mu \nu} = \chi_{\nu \mu} \,.
\end{equation}
Substituting ${R}_{\rho \sigma \mu \nu}\to \widetilde{R}_{\rho \sigma \mu \nu}$ in the original Lagrangian $\mathcal{L}$ leads to the modified Lagrangian
\begin{equation}\label{Ltilde}
	\widetilde{\mathcal{L}} \equiv \mathcal{L} \big(\widetilde{R}_{\rho \sigma \mu \nu}, g^{\alpha \beta} \hspace{0.1em} \big) \,,
\end{equation}
which is exploited to find the characteristic functions for $\mathcal{L}$. Indeed, using the chain rule together with \eq{eqn:firstderivL} and \eq{ABCdef}, and evaluating \eq{Ltilde} and its first two $\alpha$-derivatives  on a maximally symmetric background, and then setting  $\alpha$ to zero, we find %
\begin{eqnarray}
\widetilde{\mathcal{L}} \Big|_{\alpha = 0} &=&L(R)\,, \label{eqn:detL}
\\
	\label{eqn:detE}
		\ \pderiv{\widetilde{\mathcal{L}} }{\alpha} \Big|_{\alpha = 0} &=&
		 \frac{\chi (\chi - 1)}{d (d - 1)}  \, E(R)\,,\\
		\pderiv{^2\widetilde{\mathcal{L}} }{\alpha^2} \Big|_{\alpha = 0} 
		&=& \frac{\chi (\chi - 1)}{d^2 (d - 1)^2} \left[ A(R) \chi (\chi - 1) + B(R) (\chi - 1) + 2 C(R) \right] \,.
	\label{eqn:detABC}
\end{eqnarray}
We observe that for any given $\mathcal{L}$ the characteristic functions are now unambiguously determined and can   be read off conveniently from \eq{eqn:detL}, \eq{eqn:detE} and \eq{eqn:detABC}. In doing so, it can also be checked that the derivative relations \eq{eqn:firstderivrel} and \eq{eqn:secderivrel} are indeed satisfied. As such, the algorithm is highly efficient in that it circumnavigates the more tedious computation of the derivatives \eq{eqn:firstderivL} and \eq{basis}  for any given $\mathcal{L}$. We defer the sample derivation of  characteristic functions for specific models of higher order gravity to \sct{sct:compotherflow}.

It is worth noting that different curvature invariants in $\mathcal{L}$ do not necessarily result in different values for the parameters $L$, $B$, and $C$ beyond quadratic order in curvature. This is due to the fact that starting from cubic order onwards there exist many more curvature invariants than independent functions characterising the Lagrangian and its Hessian on a maximally symmetric background. Amongst others, this implies the existence of curvature invariants generating zeros for all three characteristic functions, i.e.~curvature invariants which vanish on maximally symmetric backgrounds as well as their second variation.  An example for the latter is given by the seminal Goroff-Sagnotti term $\sim C_{\rho \sigma}{}^{ \mu \nu} C_{ \mu \nu} {}^{\lambda\tau}C_{\lambda\tau}{}^{\rho \sigma}$ \cite{Goroff:1985sz,Goroff:1985th} where $C$ denotes the Weyl tensor.

\subsection{Flows for Higher Order Gravity}
\label{sct:floweqmain}
After finding the Hessians in \eq{eqn:Hessians}, we can now use \eq{fRG} to derive the flow of actions of the form \eq{eqn:bGammaform}. Generally, and even without adapting the technical choices discussed in \sct{sct:funcrenorm}, the result takes the form
\begin{equation}
	\partial_t L = \bar{I} [L, B, C] \, ,
	\label{eqn:flowdimful}
\end{equation}
where the right-hand side $\bar{I} [L, B, C] = \bar{I} [L, B, C] (R)$ arises entirely due to quantum fluctuations, and we sometimes refer to it as the fluctuation integrals. As such, $\bar I [L, B, C]$ is  the result of performing the functional traces of \eq{fRG} and a function of the background curvature. This form of the flow as a functional of $L$, $B$, and $C$ is independent of any technical choices explained in \sct{sct:funcrenorm} or the form of the regulator. It solely arises from the form of the hessian in \eq{eqn:Hessians}.
For the technical choices made in \sct{sct:funcrenorm}, in particular  \eq{eqn:regrep} and the shape function \eq{cutofffunc}, $\bar{I} [L, B, C]$ depends on curvature derivatives of $L$ as well as the flow of $L$, $B$, and $C$ due to the term $\partial_t \mathcal{R}_k$ on the right-hand side of \eq{fRG}. 

For the purpose of analysing the renormalisation group flow and finding fixed points it is  convenient to transition  from  \eq{eqn:flowdimful}  to expressions in terms of dimensionless quantities. We re-scale the background curvature in units of the RG scale $r = R/k^2$, and likewise  the functions $L$, $B$ and $C$, by writing
\begin{eqnarray}
	\ell (r) &=& L (R)/ k^{d} \,,\\ 	\label{eqn:ldimless}
	b(r) &=& B(R)/ k^{d-4} \,,\\
	c(r) &=& C(R)/ k^{d-4} \,.
\end{eqnarray}
Further, the operator traces  also depend on dimensionful technical parameters, i.e.~the endomorphism parameters $E_i$ \eq{eqn:laplaceposdef}. Since these are linear in the Ricci curvature, we  introduce their dimensionless 
counterparts $e_i$ as
\begin{equation}
	e_i = \frac{E_i}{R} \,, 
\end{equation}
which therefore are numbers bounded by  the constraints \eq{eqn:endobounds}.
In these conventions, the  flow equation takes the form
\begin{equation}
	 \partial_t \ell+d \, \ell - 2 r \ell' = I [\ell, b, c] 
	\label{floweqnabc}
\end{equation}
in general dimensions $d$, with the dependence on $e_i$ being implicit. The new terms on the left-hand side of \eq{floweqnabc} arise from the transition to dimensionless variables and account for the canonical mass dimension of the Lagrangian $[\mathcal{L}]=d$  and the mass dimension of  Ricci scalar $[R]=2$. The fluctuation integral $I$ relates to $\bar I$ in \eq{eqn:flowdimful} as
\begin{equation}
	I [\ell, b, c] (r) = k^{-d} \bar{I} [k^d \ell, k^{d - 4} b, k^{d - 4} c] (k^2 r) \,.
\end{equation}
The explicit expressions  for $I$ are rather lengthy and delegated to the App.\ref{sec:floweq}. The flow equation \eq{floweqnabc} is one of the central new result of this work. We therefore briefly discuss its general structure, and some of its basic features.
\begin{itemize}
\item[$(a)$]  {\it Structure of the flow.}\\ 
The flow equation  \eq{floweqnabc} takes the form of a non-linear partial differential equation
for the three functions $\ell$, $b$, and $c$. 
 The left-hand side shows the flow $\partial _t\ell$ and  canonical terms. The right-hand side, due to quantum fluctuations,
 can be written as
\begin{equation}
	\begin{split}
		I [\ell, b, c] =& \, I_0 [\ell, b, c] + I_1 [\ell, b, c] \partial_t \ell' + I_2 [\ell, b, c] \partial_t \ell''+ I_3 [\ell, b, c] \partial_t b + I_4 [\ell, b, c] \partial_t c \,.
	\end{split}
	\label{eqn:flowres}
\end{equation}
The  terms  $\propto \partial_t\ell'$, $\partial_t \ell''$, $\partial_t b$, and $\partial_t c$,   
are a consequence of the  regulator function $\mathcal{R}_k$ whose dependence  on $\ell$, $b$, and $c$ induces  their flow via $\partial_t \mathcal{R}_k$ in  \eq{fRG}.
The component functions $I_i$ in \eq{eqn:flowres} still depend on $\ell,b$ and $c$ and their field derivatives (see App.~\ref{sec:floweq} for explicit expressions), but no longer on flow terms. 
Depending on the choice for the action $\cal L$, the flow equation can be converted into a partial differential equation for a single or two coupled functions of background curvature (explicit examples will be given in \sct{sct:compotherflow}). Once the action contains several curvature invariants of the same mass dimension, 
additional flow equations using other background geometries can be invoked to close the system.

\item[$(b)$] {\it Expansions in powers of curvature.}\\
A useful approximation scheme consists in expanding the action
$\cal L$ in powers of curvature invariants $X_n$. Taking these as in \eq{eqn:ansatzgenfsb} with invariants of mass dimension $[X_n]=2n$ and dimensionless scale-dependent couplings  $\lambda_n= \lambda_n(t)$, we find 
\begin{equation}
\begin{aligned}
	\ell &= \sum_{n = 0}^\infty \lambda_n\, \ell_n\, r^n \, , 	\qquad \\
	b& = \sum_{n = 2}^\infty \lambda_n\, b_n\, r^{n - 2} \, , 	\qquad \\
	c &= \sum_{n = 2}^\infty \lambda_n \, c_n \, r^{n - 2} \,,
\end{aligned}
\end{equation}
where the series expansions for $b$ and $c$ follow  from the results in \sct{sct:calcparam}.  The  numerical coefficients $\ell_n$, $b_n$, and $c_n$ are unknown a priori but  determined for any given  ansatz  \eq{eqn:ansatzgenfsb}. 
Hence, the flow \eq{floweqnabc} is closed and can be resolved to give  $\partial_t \lambda_n$  for all couplings.

\item[$(c)$] {\it  Fixed points and quantum scale invariance.}\\
Fixed points are the scale-independent solutions 
$\partial_t(\ell_*,b_*,c_*) =0$, implying quantum scale invariance. 
Non-trivial UV fixed points are of particular interest as candidates for an asymptotically 
safe version of quantum gravity \cite{Weinberg:1980gg}.  At a fixed point, \eq{floweqnabc} turns into an ordinary differential equation
\begin{equation}
	 d \, \ell - 2\, r\, \ell' = I_0 [\ell, b, c] \,.
	\label{floweqnabc0}
\end{equation}
Fixed points then correspond to the well-defined, finite solutions for the functions 
$\ell _*(r)$,  $b _*(r)$,  and $c _*(r)$, or, alternatively, for the couplings $\lambda_{n,*}$. 

\item[$(d)$] {\it   Limit of classical gravity.}\\ 
In the absence of quantum fluctuations, the fluctuation integrals $I$ vanish. This leaves us with the classical flow 
$\left(\partial_t+ d -2 r\,\partial_r\right)\, \ell=0$ which integrates to 
$\ell(r,t)=r^{d/2} \cdot H\left(r e^{2t}\right)$
with $H(x)$ determined by    initial values of couplings, and no dependence on $b$ and $c$. 
We observe a Gaussian  $(\ell_*=0)$ and an infinite Gaussian $(1/\ell_*=0)$ fixed point, and
a line of classical fixed points $\ell_*\sim r^{d/2}$ (for $H=$~const) reflecting 
the classically marginal curvature invariants in $d$ dimensions.
Classical general relativity with action ${\cal L}\propto R/G_N$ and Newton's coupling $G_N=g/k^2$  
then arises through  the infinite Gaussian fixed point in the infrared limit, where $k\to 0$ and $g\to 0$ while $G_N$ is
held fixed at its observed value, and irrespective of the sign  of the cosmological constant   \cite{Falls:2014tra}.
In the presence of quantum fluctuations we have $I\neq 0$. However, quantum effects become parametrically suppressed for
\beq\label{infG}
I/\ell\to 0\quad {\rm with}\quad 1/\ell\to 0\,.
\eeq
We conclude that the limit of classical general relativity or classical higher order gravity  arises from the quantum theory through
the infinite Gaussian fixed point in the deep infrared  \cite{Falls:2014tra,Falls:2017lst,Kluth:2020bdv}.

\item[$(e)$] {\it Hyperbolic backgrounds.}\\ 
In the above, we have derived the flow for actions of the form \eq{eqn:bGammaform} on spherical backgrounds, $i.e.$~maximally symmetric backgrounds with positive curvature. The very same procedure can be carried out on backgrounds with negative curvature with the only difference that the functional traces need to be evaluated accordingly.
This can be implemented straightforwardly, leading to modifications of the corresponding heat kernel coefficients or, alternatively, spectral sums. In the context of $f(R)$ gravity this has been carried out in \cite{Falls:2016msz}.

\item[$(f)$] {\it Including matter fields.} \\ 
The flow equation can straightforwardly be extended to include  matter fields. 

\end{itemize}

We now turn to a discussion of   some more technical aspects of the flow equation, and to   new features of the flow \eq{floweqnabc} due to the wider range of admissible actions $\cal L$. 
To that end, we recall that the functions $I_i$ are  rational functions of the form
\begin{equation}
	I_i [\ell, b, c] \propto  \frac{P_i^T [\ell, b, c]}{D^T [\ell, b, c]} + \frac{P_i^S [\ell, b, c]}{D^S [\ell, b, c]} 
	+\left( \frac{P^V_{c}}{D^V_c} + \frac{P^S_{c}}{D^S_c} \right)\delta_{0i}\, ,
	\label{eqn:Iistruc}
\end{equation}
which originate from the fluctuations of the various fields contributing to   \eq{fRG}. The superscripts $T$ and $S$ denote the contributions from the tensor modes $h^T_{\mu \nu}$ and the scalar trace mode $h$, respectively. Superscripts $V$ and $S$ with an additional subscript $c$ denote contributions from the auxiliary vector and scalar modes  from ghosts and Jacobians. 
These latter terms are independent of $\ell$, $b$, and $c$, and only contribute to $I_0$. 
The numerators $P$ and the denominators $D$ are polynomials in the curvature, and linear in their arguments $\ell, b$ and $c$ or derivatives thereof.
Also, the denominators in \eq{eqn:Iistruc} are universal 
and only differ between the different York modes (explicit expressions for any $P$, $D$, and $I_i$ are summarised in App.\ref{sec:floweq}).
For the remainder, we focus on special points related to zeros of the denominators $D$, or zeros of certain numerators $P$. 
\begin{itemize}
\item[$(g)$] {\it Moveable poles.}\\
We begin with the denominators due to transverse traceless modes $D^T$ and trace modes $D^S$. These are linear functions in $\ell$, $\ell'$, $b$ and $c$, with $D^S$ additionally depending on $\ell''$, with coefficients polynomial in curvature. As such, either of these may vanish for some $r_0$. We refer to these as moveable poles to reflect that their location depends  on the form of $\cal L$ and its couplings.
   Also,  these zeros cannot be avoided in general by suitable choices of technical parameters and must be taken as part of the setup. Therefore, unless otherwise stated, we set the endomorphism parameters to their natural values
\begin{equation}\label{e1e4}
	e_1 = 0\, , \qquad e_4 = 0 \,.
\end{equation}
Explicit studies have shown that  if zeros of the denominators arise,  they are innocuous and always accompanied by zeros of the corresponding numerators, leading to  finite and well-defined solutions $\ell(r)$ for all fields.

\item[$(h)$] {\it Avoiding spurious poles.}\\
The fluctuations of the auxiliary fields  contribute to $I_0$ and have an impact on the location of fixed point solutions \eq{floweqnabc0}. Their denominators $D^V_c$ and $D^S_c$ are given by
\begin{equation}\label{denominators}
	D^V_c = 1-(e_2+\s014) r \, , \qquad D^S_c =1 -(e_3 +\s013)r \,.
\end{equation}
Once more, we observe that \eq{denominators}  can vanish for finite curvature. Also, the corresponding numerators $P_{c}^V$ and $P_{c}^V$ cannot be made to vanish simultaneously for any finite $e_3$ or $e_4$. This implies that  finite and well-defined fixed point solutions of \eq{floweqnabc0} require that these spurious poles are compensated by other terms in \eq{eqn:Iistruc}, as has been confirmed in explicit studies \cite{Falls:2014tra,Falls:2017lst}.
However, since the zeros of \eq{denominators} only depend on the technical parameters $e_2$ and $e_3$, and are otherwise independent of $\ell$, $b$ or $c$, they can  be removed   from the outset by a suitable choice of parameters
\begin{equation}\label{e2e3}
	e_2 = - \s0{1}{4} \, , \qquad e_3 = - \s0{1}{3} \,.
\end{equation}
It has been noted that the technical simplifications achieved by the choices \eq{e2e3} lead to improved results \cite{Benedetti:2012dx,Falls:2018ylp}.

\item[$(i)$] {\it Fixed singularities.}\\
The zeros of certain numerators $P$ in \eq{eqn:Iistruc} have a significance for fixed point solutions  \eq{floweqnabc0}. The reason for this is that $I_0$ is linear in $\ell$, $b$ and $c$ and their derivatives  $\ell'$, $\ell''$,  $\ell'''$, $b'$ and $c'$ with polynomial coefficients in curvature. To illustrate our points, we first consider theories with $b=c=0$, relevant for $f(R)$ type models of quantum gravity. In this case,  the trace modes $h$ generate a term $\sim  \ell'''$, with
\beq\label{ell'''}
P_0^S[\ell, 0,0] =P_0^{S\ell3} \ell''' + \cdots \,,
\eeq
and  dots indicating further terms with lower derivatives of $\ell$, see \eq{PS}. The coefficient $P_0^{S\ell3}(r,e_4)$, given in \eq{lppprefac}, is a quintic polynomial in $r$ without constant term which further depends on $e_4$.
The fixed point condition \eq{floweqnabc0} then becomes a third order differential equation for the fixed point function $\ell$. 
Expressed in normal form, it becomes an ordinary  third order non-linear differential equation
\begin{equation}
	\ell''' = {\mathcal{J} (\ell, \ell', \ell'', r)}/{P_0^{S \ell 3}(r)} 
	\label{eqn:lpppfr}
\end{equation}
for some function $\mathcal{J}$, and provided that $P_0^{S \ell 3}\neq 0$. Background curvatures $r=r_0$ where $P_0^{S \ell 3}$ vanishes are referred to as singularities, and take a special role in that they change the order of the differential equation. 
In general, one zero of $P_0^{S \ell 3}$ is always located at $r_0 = 0$. 
In addition, we always find two real and a  complex conjugate pair of solutions $r_0$ for any value of $e_4$. 
For example, for vanishing endomorphism parameter $e_4$, the fixed singularities are located at
\begin{equation}
\begin{split}
	r_0 &\approx -9.9986 \,,\qquad \\
	r_0 &= 0 \,, \qquad \\
	r _0&\approx 2.0065 \,,  \qquad \\
	r_0 &\approx -4.9763 \pm 0.46851 i \,.
\end{split}
\end{equation}
Also, in contrast to the spurious poles from the auxiliary sector it is not possible to remove these zeros by an appropriate choice for $e_4$.
Hence,  for solutions of \eq{floweqnabc0}   to remain well-defined even across $r_0$, a compensating zero of $\mathcal{J}$ is required for any  zeros of $P_0^{S \ell 3}$  along the real axis in field space.
This transforms the search for global fixed points into a boundary value problem for \eq{eqn:lpppfr}: each possible singularity requires the fine-tuning of one  open parameter of the general solution  to ensure that $\ell$ remains well-defined for all real $r$. In particular, if the number of zeros of $P_0^{S \ell 3}$ is equal to the order of the differential equation, only a countable number of well-behaved  solutions may exist.\footnote{Examples where this has been carried out for $f(R)$ gravity include \cite{Dietz:2012ic,Demmel:2014sga,Demmel:2015oqa}.} For models of quantum gravity with $b=c=0$ we conclude that the trace-mode-induced coefficient $P_0^{S \ell 3}$   has a direct impact on the possible space of fixed point solutions.

\item[$(j)$] {\it Avoiding fixed singularities.}\\
New features arise if  actions \eq{eqn:bGammaform}  are permitted with either $b$ or $c$ or both different from zero.
We illustrate our point, exemplarily,  for models where $b$ is proportional to $\ell''$, and $c=0$. 
Owing to  \eq{eqn:ghostdof} and \eq{eqn:scadof}, these higher order models of gravity display additional propagating spin-2 degrees of freedom, and, possibly, additional massive spin-0 degree degrees of freedom. Once more, interacting fixed point solutions of  \eq{floweqnabc0}   arise as a third order differential equation for $\ell$. 
Terms proportional to $\ell'''$ continue to be generated by the fluctuations of the $h$ modes. In contrast to the previous example, however, additional contributions arise through the transverse traceless modes $h_{\mu \nu}^T$. This is so because
\begin{equation}\label{b'}
		P_0^T[\ell,b,0] =P_0^{T b1} \,b' + \cdots 
\end{equation}
see \eq{PT}, with $b'\propto \ell'''$ and dots indicating terms involving  lower $\ell$-derivatives. The coefficient $P_0^{T b1}$ is an $e_2$-dependent quintic polynomial in curvature without constant term, see \eq{eqn:P0Tb1}.
Bringing  \eq{floweqnabc0}  with  \eq{ell'''} and \eq{b'} into normal form, we find
\begin{equation}
	\ell''' = \frac{\mathcal{K} (\ell, \ell', \ell'', r)}{D^T [\ell,b,0] P_0^{S \ell 3}  + D^S [\ell,b,0] P_0^{T b 1}} 
	\label{eqn:normformtt}
\end{equation}
for some function $\mathcal{K}$ and with $P_0^{S b1}$ as in \eq{eqn:lpppfr}. 
The fact that the transverse traceless modes also generate a term $\propto \ell'''$ changes the nature of the fixed point differential equation. 
Comparing \eq{eqn:lpppfr}  with \eq{eqn:normformtt} we observe that the denominator of  \eq{eqn:normformtt}
now additionally depends on $D^T$ and $D^S$, and hence on the couplings of the theory through $\ell$ and $b\propto \ell''$. 
Ultimately, this is due to the trace and transverse modes carrying different denominators  \eq{eqn:Iistruc}. 
Most notably, unlike in  \eq{eqn:lpppfr} where the singularities are fixed, the singularities of  \eq{eqn:normformtt} have been rendered movable owing to the higher order nature of the underlying models. Hence, in these more general setups, the quantum dynamics of the theory itself determines whether and where singularities due to vanishing denominators in \eq{eqn:normformtt}  arise, if at all.
A  more detailed quantitative analysis of this aspect in higher order theories of quantum gravity is deferred to  a forthcoming publication \cite{forthcoming}.
\end{itemize}
This concludes the discussion of general features of the flow equation \eq{floweqnabc}  for higher order theories of gravity with fundamental actions \eq{eqn:bGammaform}, and the condition for interacting fixed points \eq{floweqnabc0}.

\section{\bf Applications}
\label{sct:compotherflow}
In this section we explain how our setup can be used to study the effects of different higher order curvature invariants. We show how operator traces on maximally symmetric backgrounds can be used to project the flow onto specific curvature monomials. We also revisit flow equations for template models of quantum gravity  studied in the literature and derive their characteristic functions $L$, $E$, $A$, $B$, and $C$.

\subsection{General Projections}
\label{sct:projection}
Flow equations on maximally symmetric backgrounds are particularly useful when considering (derivative) expansions of the quantum effective action $\bGamma_k$ which contain a single operator for each mass dimension, see \eq{eqn:ansatzgenfsb}.
Let us now  discuss how the functional renormalisation group generates the flow for actions of the form \eq{eqn:ansatzgenfsb} using a general background geometry before specialising to a maximally symmetric background and discussing which approximation are implied by that.
In the present setup \eq{eqn:wetterichbga}, the flow equation for actions \eq{eqn:ansatzgenfsb} generate a sum of operators $\{X_n\}$ on the left-hand side. On the right-hand side, the operator trace generates all possible curvature monomials, including some which are not part of the set  $\{X_n\}$ retained in the initial action.  To make this more explicit, we introduce a complete basis of curvature invariants $\{Y_{n, i}\}$ with $n$ labelling the mass dimension, as before, and $i$ labelling the different operators of equal mass dimension. Without loss of generality we can choose this new basis such that $Y_{n, 1} = X_n$. After computing the functional traces, \eq{eqn:wetterichbga} can be written into the form
\begin{equation}\label{project}
	\sum_{n = 0}^\infty \int \text{d}^d x \sqrt{g} \, \overline{\beta}_n X_n = \sum_{n = 0}^\infty \sum_i \int \text{d}^d x \sqrt{g} \, \zeta_{n, i} Y_{n, i} \, ,
\end{equation}
where $\overline{\beta}_n$ are the dimensionful $\beta$-functions of the couplings $\blambda_n$ and $\zeta_{n, i}$ are functions depending on the couplings and potentially their $\beta$-functions. 

Since we are only interested in the flow of $\blambda_n$ associated to the operators $X_n$, we require a  procedure to project the right-hand side onto the operators $X_n$. It is important to note that this projection is generally ambiguous due to the absence of a natural scalar product between different curvature invariants. Hence, the projection will depend on the chosen basis for the curvature monomials. After constructing a complete basis $\{Y_{n, i}\}$, the canonical projection is given by demanding that all  $Y_{n, i>1}$ in \eq{project} vanish. 
Following this projection procedure, the use of maximally symmetric backgrounds is equivalent to considering a canonical operator basis $\{Y_{n, i}\}$ in which all operators except $Y_{n, 1} = X_n$ vanish on the  chosen background, $Y_{n, i>1}|_{\rm msb}=0$. Then, our projection procedure of setting $Y_{n, i>1} = 0$  is equivalent to evaluating all operators on the background geometry. 

More generally, starting from an arbitrary operator basis $\{Z_{n, i}\}$, a canonical basis $\{Y_{n, i}\}$ can always be constructed provided the curvature monomials $X_n$ are non-vanishing on the background. That this is always possible  can be appreciated by expressing all curvature invariants using the Ricci decomposition, whereby the
Ricci scalar curvature $R$, the traceless Ricci tensor $S_{\mu \nu}$ and the Weyl tensor $C_{\rho \sigma \mu \nu}$ 
\beq
\{R, S_{\mu \nu}, C_{\rho \sigma \mu \nu}\}
\eeq
are used as building blocks to construct any curvature invariant of mass dimension $2n$ for any positive integer $n$. Then, any operator in the basis $\{Z_{n, i}\}$ takes the form
\begin{equation}
	Z_{n, i} = z_{n, i} R^n + \mathcal{O} \left( S_{\mu \nu}, C_{\rho \sigma \mu \nu} \right) \, ,
\end{equation}
with $z_{n, i}$ a possibly dimension-dependent constant, and $\mathcal{O} \left( S_{\mu \nu}, C_{\rho \sigma \mu \nu} \right)$ a sum of terms vanishing on a maximally symmetric background, i.e.~terms containing at least one power of $S_{\mu \nu}$ or $C_{\rho \sigma \mu \nu}$. Assuming without loss of generality that $X_n\equiv Z_{n, 1} = Y_{n, 1} $, the canonical basis $\{Y_{n, i}\}$ is explicitly given by
\begin{equation}
	 X_n\equiv Y_{n, 1} = Z_{n, 1}  \, , \qquad Y_{n, i} = Z_{n, i} - \frac{z_{n, i}}{z_{n, 1}} Z_{n, 1} \quad \forall \quad i \neq 1 \,.
\end{equation}
By construction, all $Y_{n, i>1}$  vanish on the background geometry as required. Note that the only requirement on $X_n$ is that it is non-vanishing on a maximally symmetric background, 
\begin{equation}
	X_n \equiv Z_{n, 1} = z_{n, 1} R^n + \mathcal{O}\left( S_{\mu \nu}, C_{\rho \sigma \mu \nu} \right) \,,
	\label{eqn:curvnonvanfsb}
\end{equation}
with $z_{n, 1} \neq 0$. 
An example where the $X_n$ do not receive any contributions from $S_{\mu \nu}$ and $C_{\rho \sigma \mu \nu}$ is given by powers of the Ricci scalar $X_n\sim R^n$, and corresponds to a projection onto the curvature monomials contained in $f(R)$ models of gravity. However, we are not limited to this case and may also project onto curvature invariants containing $S_{\mu \nu}$ and $C_{\rho \sigma \mu \nu}$ as long as $z_{n, 1}$ does not vanish. Examples for the latter  have been studied in  \cite{Falls:2017lst,Kluth:2020bdv}. 
 
Finally, our discussion also highlights well-known limitations of maximally symmetric backgrounds. As soon as the decomposition of a curvature monomial $X_n$ as in \eq{eqn:curvnonvanfsb} has no term $\sim R^n$, its flow on maximally symmetric backgrounds cannot be determined.
Further, as indicated above, maximally symmetric backgrounds constrain the types of curvature bases that can be used for the projection. In particular, non-canonical curvature bases $\{Y_{n, i} \}$ where some  $Y_{n, i>1} |_{\textrm msb} \neq 0$ necessitate additional input, $eg.$~less symmetric background geometries, to disentangle the flow of couplings. 
The latter equally applies if several  field monomials of the same canonical mass dimension are retained. In the remainder of this section, we discuss various examples of increasing complexity, and explain how the flow for general curvature invariants of the form \eq{eqn:curvnonvanfsb} can be analysed within our framework.

\subsection{Einstein-Hilbert}
As a first example, we discuss  the renormalisation group flow for the Einstein-Hilbert action  which has been studied in many incarnations of the functional RG, e.g.~\cite{Reuter:1996cp,Souma:1999at,Souma:2000vs,Lauscher:2001ya,Reuter:2001ag,Litim:2003vp}. Here, we have
\begin{equation}
	\mathcal{L}  = \blambda_0 + \blambda_1 R \, ,
\end{equation}
with $\blambda_n$ the dimensionful couplings. Since this Lagrangian does not include any operators quadratic in curvature, it follows straightforwardly that
\begin{equation}\label{EHabc}
\begin{split}
	L &= \blambda_0 + \blambda_1 R \, , \qquad \\
	E &= \blambda_1 \,,\qquad\\
	A &= B = C = 0 \, .
\end{split}\end{equation}
We introduce dimensionless couplings
\begin{equation}
\begin{split}
	\lambda_0 &=16 \pi \,{\blambda_0}/{k^4}  \,, \\
	 \lambda_1 &= 16 \pi\,{ \blambda_1}/{k^2} \,,
\end{split}
\end{equation}
where the factor of $16 \pi$ is purely conventional and chosen such that  Newton coupling in units of the RG scale $g=G_k/k^2$ is given by minus the inverse of $\lambda_1$. Also using the dimensionless Ricci curvature $r=R/k^2$, we obtain the beta functions $\beta_i$ from \eq{floweqnabc} with \eq{EHabc}. Neglecting all terms quadratic or higher in $r$, we find
\begin{equation}\label{EHgeneral}
	\begin{split}
		\beta _0  + 4 \lambda _0 + r \left( \beta _1 + 2 \lambda _1 \right) =& \frac{1}{24 \pi} \bigg[ \, \frac{30 \left(\beta _1 (6 e_1 r + r - 2)+2 \lambda _1 (18 e_1 r + 3 r-8)\right)}{\lambda _1 (6 e_1 r + (3 \tau - 4) r - 6)+6 (\tau - 1) \lambda _0} \\
		& \qquad - \frac{3 \beta _1 (6 e_4 r - r - 2)+6 \lambda _1 (18 e_4 r - 3 r - 8)}{2 (1 + \tau) \lambda _0 + (3 + r \tau - 3 e_4 r) \lambda _1} \\
		& \qquad+ 36 e_2 r + 12 e_3 r - 23 r - 48 \bigg] \,,
   	\end{split}
\end{equation}
which can be resolved for $\beta_i$. 
To make a link with  the notation of \cite{Reuter:1996cp} we express the action in terms of the cosmological 
constant $\lambda$ and Newton's coupling $g$, which  are related to $\lambda_0$ and $\lambda_1$ by
\begin{equation}
\begin{split}
	\lambda &= - {\lambda_0}/(2 \lambda_1) \,,\\ 
	g &= - {1}/{\lambda_1} \, .
\end{split}
	\label{eqn:lambdag}
\end{equation}
Then, expanding the denominators in small curvature, exemplarily   for vanishing endomorphisms and $\tau = 0$, we find 
\begin{equation}\label{EH0}
	\begin{split}
		\partial_t \lambda =& \, \lambda (\eta - 2) + \frac{\lambda (42 - 96 \lambda + 13 \eta) -9 (\eta - 2)}{12 \pi (2 \lambda - 1) ( 4 \lambda - 3)} g \, , \\
		\partial_t g =& \, (2 + \eta)g \, , \\
		\eta =& \, 3 g \frac{237 -680 \lambda+ 756 \lambda^2 - 368 \lambda^3}{72 \pi (1 - 2\lambda)^2 (4 \lambda -3) + 2 g (48 -97 \lambda + 42 \lambda^2)} \, ,
	\end{split}
\end{equation}
where $\eta$ is the anomalous dimension of the graviton. 
The flow \eq{EH0} features the well-known Reuter fixed point \cite{Reuter:1996cp} located at
\begin{equation}
	\label{eqn:ReuterFP}
\begin{split}
	\lambda &= 0.12926 \,,\\
	 g &= 0.98416 \,,
\end{split}
\end{equation}
with critical exponents
\begin{equation}
	\label{eqn:ReuterEigV}
	\theta_{0/1} = -2.3824 \pm 2.1682 i \, ,
\end{equation}
in agreement with the results in \cite{Falls:2017lst,Falls:2014tra}. For non-trivial choice of the endomorphism parameters $e_2$ and $e_3$ we also recover \cite{Falls:2018ylp,Kluth:2020bdv} with small changes in the numerical values for the couplings and the eigenvalues compared to \eq{eqn:ReuterFP} and \eq{eqn:ReuterEigV}.

\bfi
	\includegraphics[width=.45\linewidth]{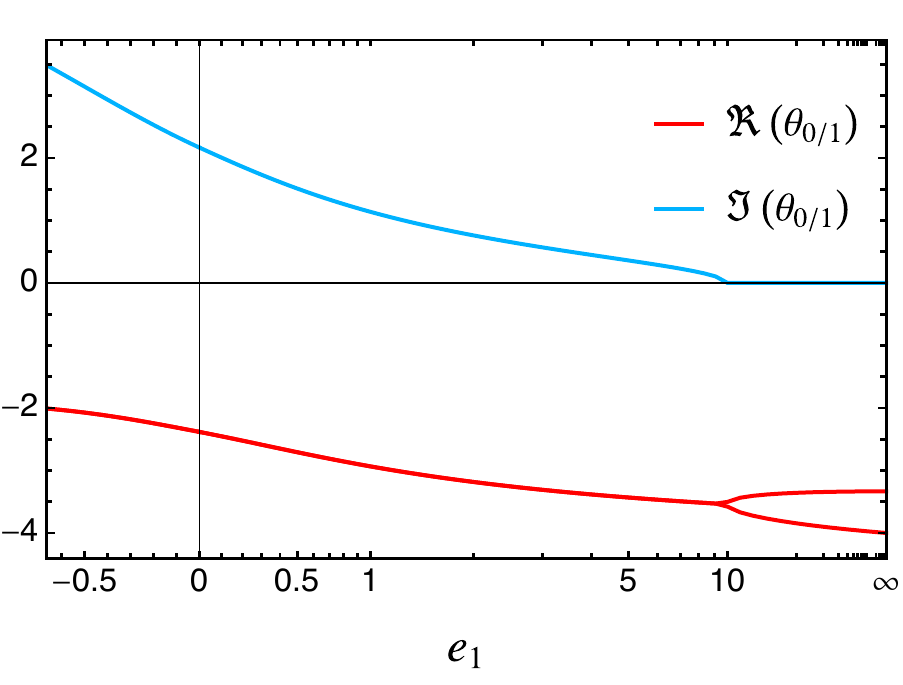}
	\caption{Shown is the dependence of the eigenvalues of the Reuter fixed point on the endomorphism parameter $e_1$. In line with the bounds given in \eq{eqn:endobounds}, we present the whole range of allowed values for $e_1$ while all other parameters have been set to zero, i.e. $\tau = e_2 = e_3 = e_4 = 0$. The red line indicates the real part of the eigenvalues which are complex conjugated until $e_1 \approx 10$. The absolute value of the imaginary part for both eigenvalues is displayed by the blue line.}
		\label{fig:endotau}
\efi

\bfi
	\includegraphics[width=.45\linewidth]{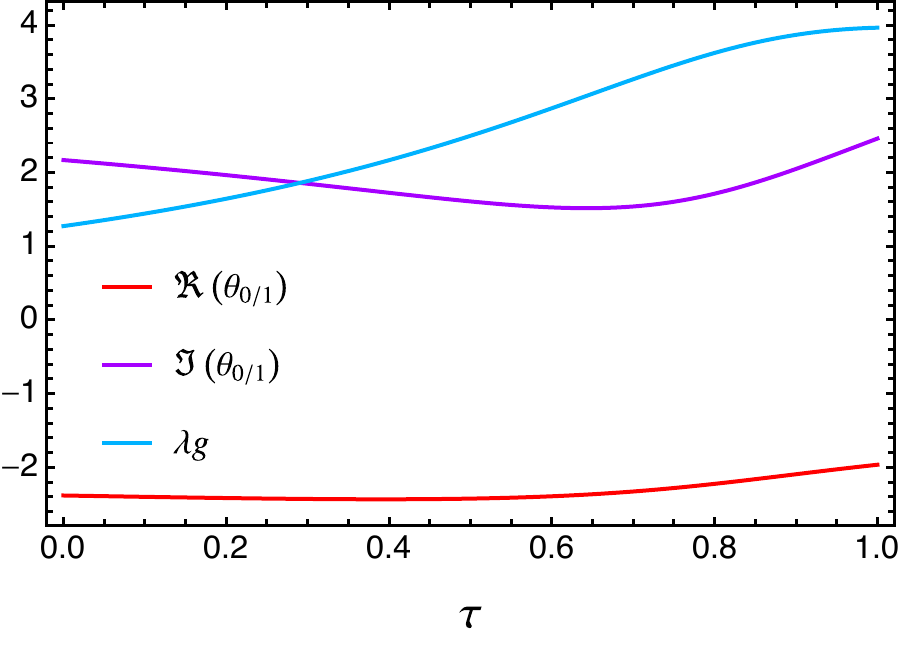}
	\caption{
	Shown are the scaling exponents and the product of couplings $\lambda g$ at the Reuter fixed point, and their dependence on 
	the metric split parameter $\tau$, \eq{eqn:taumetricsplit}, also using $e_i = 0$. 
	Interpolating between linear $(\tau = 0)$ and  exponential split $(\tau = 1)$, the $\tau$ dependence of eigenvalues is 
	 mild.}
		\label{fig:tau}
\efi

\bfi
	\includegraphics[width=.5\linewidth]{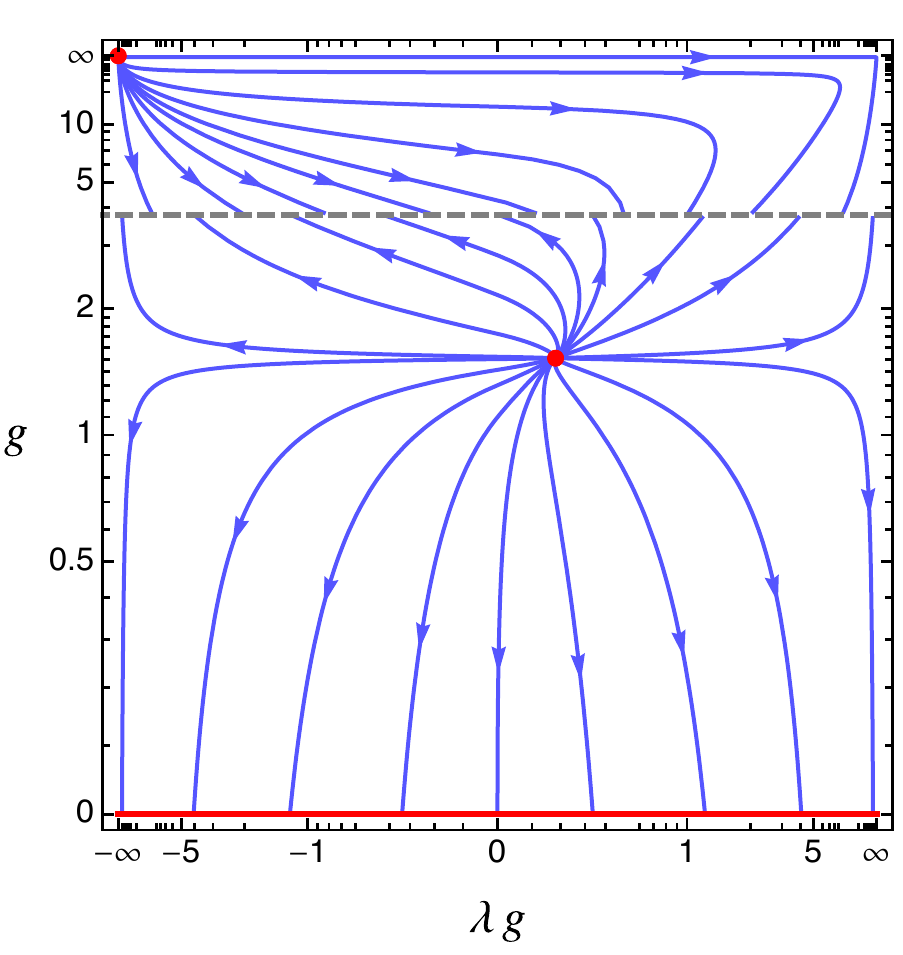}
	\caption{The phase diagram of Einstein-Hilbert gravity in terms of $\lambda$ and $g$ with arrows on  trajectories pointing towards the infrared. The grey dashed line indicates a singularity of the anomalous dimension and separates regimes of weak and strong coupling.  The lower panel shows the Reuter fixed point (central red dot), and trajectories connecting it with classical general relativity in the infrared (red horizontal axis, $g=0$). The upper panel shows a strongly coupled UV fixed point $1/g=0=1/(g\lambda)$ (red dot). Trajectories emanating from the latter terminate at the singularity of the anomalous dimension and cannot reach classical general relativity. 
	}
	\label{fig:phase}
\efi

In \fig{fig:endotau}, we show the dependence of the Reuter fixed point on $e_1$ over the whole range of parameters in accordance with \eq{eqn:endobounds} while the remaining endomorphisms and $\tau$ fixed to $0$. Both eigenvalues are relevant throughout. We find  complex conjugate pairs  for small ($e_1 \lesssim 10$), and real eigenvalues for large endomorphism. In the latter, couplings  $\lambda$ and $g$ scale as $1/e_1$.

In \fig{fig:tau}, we show the eigenvalues of the fixed point and $\lambda g$ in an interpolation between the linear and the exponential split while keeping all endomorphism parameters vanishing. While the eigenvalues never become real we observe that they stay relevant in the whole range giving qualitatively the same result in the linear and the exponential split. Note that this correspondence between the linear and the exponential split seems to hold in the Einstein-Hilbert approximation while it has been observed that higher curvature invariants such as in an $f(R)$ expansion modify this result \cite{deBrito:2018jxt}.

\subsection{Phase Diagram}

The phase diagram of Einstein-Hilbert gravity has been studied in many works, e.g.~\cite{Reuter:2001ag, Litim:2003vp,Fischer:2006fz,Donkin:2012ud,Christiansen:2012rx,Gies:2015tca}.
Here, we exploit the  flow equation \eq{EHgeneral} to find the phase diagram of Einstein-Hilbert gravity
and general analytical solutions for the running of couplings. This benefits from the limit
of large endomorphism $1/e_1 \rightarrow 0$, also using suitably rescaled couplings  $\lambda \rightarrow \frac{\lambda}{e_1}$ 
and $ g \rightarrow \frac{g}{e_1}$. In this limit, the $\tau$-dependence drops out. Another feature is that  the flow for the 
cosmological constant decouples naturally (Fig.~\ref{fig:endotau}). The resulting $\beta$-functions take simple analytical 
expressions   given by
\begin{equation}
	\begin{split}
		\partial_t \lambda &=    \left( {\lambda} - \frac{\lambda_*}{g_*}g \right) 
		\frac{\theta_0 - {\theta_1}\,\left(1-g_*/g\right)}{1+  \s012\,\theta_1\,(1-g_*/g)}\ , \\[1ex]
		\partial_t g &= 
		\theta_1\frac{g-g_*}{1+\s012(1-g_*/g)\theta_1}\,.
	\end{split}
	\label{eqn:infendobetas}
\end{equation}
Notice that \eq{eqn:infendobetas} only depends on the fixed point coordinates and eigenvalues, which in our setup are given by
\begin{equation}
	\lambda_* = \frac{3}{25} \, , \qquad g_* = \frac{12 \pi}{25} \, , \qquad 
	\theta_0 = -4 \, , \qquad \theta_1 = - \frac{10}{3} \, ,
	\label{eqn:fpinfendo}
\end{equation}
in accord with  \fig{fig:endotau}. We observe that $\lambda$ no longer couples into the flow of $g$, giving rise to the eigenvalue \eq{CCscaling}.
The flow for $g$ can also be written in terms of the graviton anomalous dimension $\eta$
\begin{equation}
	\begin{split}
		\partial_t g =& \, (2 + \eta)\, g\,,\quad{\rm where}\quad  \eta=-\frac{2}{ 1+ \s012 \theta_1(1-g_*/g)} \, .
	\end{split}
\end{equation}
The explicit expressions make it evident that the anomalous dimension $\eta$ approaches $-2$ at  the UV fixed point. 
Also, the couplings approach the UV with scaling exponents $\theta_0$ and $\theta_1$, respectively. 
Notice that the anomalous dimension diverges for $g\to g_{\rm bound}=\frac{\theta_1}{2+\theta_1}\,g_*$, with $ g_{\rm bound}>g_*$ for $\theta_1<-2$.

The flow $\partial_t g$ in \eq{eqn:infendobetas} can be integrated in closed form since it is independent of the cosmological constant \cite{Litim:2003vp,Litim:2007iu,Gerwick:2011jw}. Then, for $g$ away from its fixed point and $g_{\rm bound}$, we find that the running is monotonous, 
\begin{equation}
	\left( \frac{g}{g_0} \right)^{\tfrac{1}{2}} \left( \frac{g_* - g}{g_* - g_0} \right)^{\tfrac{1}{\theta_1}} = \frac{k}{k_0} \, ,
	\label{eqn:infendoganalytic}
\end{equation}
with $g_*$ and $\theta_1$ as defined in \eq{eqn:fpinfendo}. The powers of the first and second factor of \eq{eqn:infendoganalytic} 
relate to the inverse scaling exponent of $g$  at the infrared fixed point ($\theta_{\rm IR} = 2$) and the ultraviolet fixed point ($\theta_{\rm UV} = \theta_1$), respectively. From these expressions
 we can easily read off the Gaussian and the Reuter fixed point. 

For $g\neq g_*$, we  also find the analytical solution of $\lambda$ as a function of the running Newton's coupling \eq{eqn:infendoganalytic},
\begin{equation}
	\lambda =  \lambda_h\left(\frac{g}{g_*}\right)+\left[ \lambda_0  - \lambda_h\left(\frac{g_0}{g_*}\right)\right]\, 
	\frac{g_0}{g}\,\left( \frac{g_* - g}{g_* - g_0} \right)^{\tfrac{\theta_0}{\theta_1}} \,.
	\label{eqn:cosmolambdaanalytic}
\end{equation}
where we have introduced the auxiliary function
\begin{equation}\label{aux}
	\lambda_h(x) = \frac{\lambda_*}{{\theta_0} - 2\theta_1} \left[ x\,  ({\theta_0 - \theta_1})
	- \frac{\theta_0\,\theta_1}{\theta_0 - \theta_1} + \frac1x \frac{(\theta_1)^2}{\theta_0-\theta_1} 
	\right]\,,
\end{equation}
In the high energy limit $g\to g_*$ we observe  $\lambda\to \lambda_*=\lambda_h(1)$, in agreement with \eq{eqn:fpinfendo}. 
For $g=g_*$, the running of $\lambda$ is independent of $g$ and reads $\lambda=\lambda_*-(\lambda_*-\lambda_0)e^{t \theta_0}$. 
Simple analytical expressions  for the running of couplings such as \eq{eqn:infendoganalytic}, \eq{eqn:cosmolambdaanalytic} and \eq{aux} are  useful  for many farther reaching applications in particle physics, cosmology or black holes
\cite{Litim:2007iu,Falls:2010he,Gerwick:2011jw,Hindmarsh:2011hx,Falls:2012nd,Litim:2013gga}.

\fig{fig:phase} shows the phase diagram for Einstein-Hilbert gravity in terms of $\lambda$ and $g$. The resulting plot shows trajectories connecting the Reuter fixed point in the ultraviolet with the  fixed point $g\to 0$ in the infrared.\footnote{The fixed point $g=0$  corresponds to the infinite Gaussian fixed point $1/\ell\to 0$ discussed in Sect.~\ref{sct:floweqmain}.} As such, we observe that low energy regimes with positive, negative, or vanishing cosmological constant can be achieved. The phase diagram also displays a boundary in the strong coupling domain 
 at  $g_{\rm bound} =  \tfrac{6 \pi}{5} \approx 3.77>g_*$ where the anomalous dimension in \eq{eqn:infendobetas} becomes singular, with $\eta<0$ ($\eta>0$) below (above) the boundary. Trajectories running into the boundary terminate. Along the boundary  the resulting $\beta$-functions are ill-defined, and we find that $g$ becomes complex by using the full analytic solution in \eq{eqn:infendoganalytic}. Also, above the boundary, RG flows are once more well-defined. In this regime, we find a strongly interacting fixed point at 
$1/g_*=0$ and $1/(g\lambda)_*=0$ with $\eta_*=3$. This strong coupling fixed point is ultraviolet with two relevant eigendirections. However,
all emanating trajectories terminate at $g_{\rm bound}$,  and cannot reach the low energy regime where classical general relativity becomes valid.

\subsection{Gauss-Bonnet}

Next, we consider Gauss-Bonnet gravity which we take to be Einstein-Hilbert gravity amended by  the Gauss-Bonnet term
\begin{equation}
\begin{split}
	\mathcal{L} &= \blambda_0 + \blambda_1 R + \blambda_{\rm GB} \,{\rm GB} \,,\\
	{\rm GB} &= R^2 - 4 R_{\mu \nu} R^{\mu \nu} + R_{\rho \sigma \mu \nu} R^{\rho \sigma \mu \nu} \,.
\end{split}
\end{equation}
The Gauss-Bonnet term fulfils
\begin{equation}
	\int \text{d}^4 x \sqrt{g} \, {\rm GB} = 32 \pi^2 \chi (M) 
\end{equation}
in four dimensional spacetime with $\chi(M)$ the Euler characteristic. On a spherical background we find ${\rm GB} = R^2/6$ leading to the familiar result
\begin{equation}
	\chi (M) = 2 \, .
\end{equation}
Since the Gauss-Bonnet term is a topological invariant in four dimensional spacetime, its first variation  is a total derivative and, therefore, it cannot contribute to Hessians of the action. Still, the parameters $L, E, A, B$, and $C$ are non-vanishing
\begin{equation}\label{eqn:paramGB}
\begin{split}
	L &= \blambda_0 + \blambda_1 R + \s016 \blambda_{\rm GB} \,{R^2} \, , \quad\\
	 E &= \s013 \blambda_{\rm GB} \,R\,,\quad\\
	A &= 2 \blambda_{\rm GB} \, , 	\quad \\
	B &= -8 \blambda_{\rm GB} \, , \quad \\
	C &= 2 \blambda_{\rm GB} \,.
\end{split}
\end{equation}
Despite of this, all terms in the Hessians
originating from the Gauss-Bonnet term  vanish due to cancellations. 
This identifies the Gauss-Bonnet coupling as an inessential one and guarantees that the flow of $\blambda_0$ and $\blambda_1$ is identical to their flow in the Einstein-Hilbert theory without Gauss-Bonnet term. 
The flow of the Gauss-Bonnet coupling (using $\lambda_{\rm GB} = 16 \pi \blambda_{\rm GB})$ is given by
\begin{equation}\label{eqn:GBeq}
	\begin{split}
		\partial_t \lambda_{\rm GB} =
				& \, \frac{g \left(482160 \lambda ^4-1155004 \lambda ^3+529036 \lambda ^2+358587 \lambda -231912\right)}{720 \pi (4 \lambda -3)(2 \lambda -1) [g(\lambda  (42 \lambda -97)+48)+36 \pi  (4 \lambda -3) (1-2 \lambda )^2]}
				\\& +\frac{13504 \lambda ^4-30692 \lambda ^3+30572 \lambda ^2-20305 \lambda
				 +6702}{5(2 \lambda -1) [g(\lambda  (42 \lambda -97)+48)+36 \pi  (4 \lambda -3) (1-2 \lambda )^2]}
\end{split}
\end{equation}
with $\lambda$ and $g$ defined as in \eq{eqn:lambdag}. 
Due to the absence of $\lambda_{\rm GB}$ in all $\beta$-functions a fixed point for $\lambda_{\rm GB}$ can only be found if the fixed point of the other $\beta$-functions induce a vanishing $\partial_t \lambda_{\rm GB}$ by chance. Otherwise, this coupling grows to plus or minus infinity. Redefining the coupling according to $\rho = 1/\lambda_{\rm GB}$ the only fixed point for $\rho$ is the asymptotically free one at $\rho = 0$.
Finally, it is worth pointing out that the independence of all other $\beta$-function on  $\lambda_{\rm GB}$  is not a result of our approximation or the choice of background geometry. Rather, this is entirely due to the topological nature of the Gauss-Bonnet term, which in turn makes the coupling an inessential one. A similar observation has been made based on studies up to quadratic order in curvature \cite{Falls:2020qhj,Knorr:2021slg}.

\subsection{${f(R)}$ Gravity}\label{fRsec}
A well known example for a gravitational action containing arbitrary high curvature invariants is given by $f(R)$ gravity with actions of the form
\begin{align}\label{fR}
	\mathcal{L} &=  f (R)  \,.
\end{align}
These types of theories have extensively been analysed in the asymptotic safety literature 
(see e.g.~\cite{Machado:2007ea,
  Codello:2008vh,
Dietz:2012ic,  
Falls:2013bv, 
%  Ohta:2013uca,
 Benedetti:2013jk,
 Dietz:2013sba,
  Falls:2014tra,
%Saltas:2014cta,
 %  Eichhorn:2015bna,
  Ohta:2015efa,
%  Ohta:2015fcu,
Demmel:2015oqa,
Falls:2016wsa,
  Falls:2016msz,
  Falls:2017lst,
  Falls:2018ylp}).
%  ,deBrito:2018jxt}. 
  These types of theories  are also contained in the general setup \eq{floweqnabc}.
To obtain the corresponding parameters, we use the results of Sect.~\ref{sct:calcparam}, and start by noting that the form of the modified Ricci scalar curvature  \eq{eqn:Riemannchi} is given by
\begin{equation}
	\widetilde{R} = g^{\rho \mu} g^{\sigma \nu} \widetilde{R}_{\rho \sigma \mu \nu} = R + \alpha\,\frac{\chi }{d}\frac{\chi - 1}{d - 1} \, .
\end{equation}
Hence, substituting the Lagrangian \eq{fR} by $\mathcal{\widetilde{L}} =  f (\widetilde{R})$   we find
\begin{equation}\label{ftilde}
\begin{split}
	\partial_\alpha \mathcal{\widetilde{L}}
	\big|_{\alpha = 0} &= f'(R) \frac{\chi (\chi - 1)}{d (d - 1)} \, , 
	\qquad \\[1ex]
	\partial^2_\alpha \mathcal{\widetilde{L}}\big|_{\alpha = 0} &= f''(R) \frac{\chi^2 (\chi - 1)^2}{d^2 (d - 1)^2} \,,
\end{split}
\end{equation}
and comparison of  \eq{ftilde} with \eq{eqn:detE} and \eq{eqn:detABC} gives
\begin{equation}
\begin{split}
	L& = f(R) \, , \qquad \\
	E &= f' (R) \, , \qquad \\
	A &= f'' (R) \, , \qquad \\
	B &= 0 \, , \qquad \\
	C &= 0 \,,
\end{split}
	\label{eqn:fRparam}
\end{equation}
confirming that $f(R)$ gravities have vanishing $B$ and $ C$. As discussed in \sct{sct:hessiansb}, this has the effect that the $\nabla^4$ term in the Hessian for the transverse traceless tensor modes are absent.
Combining \eq{eqn:fRparam} into \eq{floweqnabc} we find a general flow equation for $f(R)$ gravity with open endomorphisms and unspecified $\tau$. The choice for these parameters can have crucial effects on the type of fixed point solutions. In particular, it has been noted that solutions to the equations of motion are absent for the linear split with trivial endomorphisms parameters  $ e_i = 0$ \cite{Dietz:2012ic,Dietz:2013sba,Falls:2016wsa}. In \cite{Dietz:2013sba} it was argued that eigenperturbations of such fixed points not admitting solutions to the equations of motion are redundant by non-trivial field redefinitions. However, it turns out that solutions to the equations of motion do exist for the linear split with endomorphism parameters \eq{e1e4} and \eq{e2e3} 
\cite{Falls:2018ylp}. The latter choice also removes technical poles in the flow equation obtained from the denominators $D_c^V$ and $D_c^S$ in \eq{eqn:techpolesdenom}.

While the flow equation \eq{floweqnabc} agrees using \eq{eqn:fRparam} to some results in the literature \cite{Codello:2008vh,Dietz:2012ic,Falls:2014tra,Falls:2018ylp}, flow equations using other technical choices as explained in \sct{sct:funcrenorm} cannot or only partly be obtained from our result. In particular, note that \eq{floweqnabc} is subject to the Landau gauge with $\delta = 0$ which makes it different from flows using the unimodular gauge \cite{Alkofer:2018fxj,Ohta:2015efa,Ohta:2015fcu}. As explained above, in this gauge the physical fluctuations $\bGamma_k^{\sigma \sigma}$ enter the flow \eq{eqn:wetterichbga} rather than $\bGamma_k^{h h}$. Due to this, only the transverse tensor sector of our flow equation, i.e. $P^T$ and $D^T$ given in App.\ref{sec:floweq} agree with those works. Moreover, literature results can differ due to different techniques in evaluating functional traces in particular by using smoothed spectral sums \cite{Benedetti:2012dx,Alkofer:2018fxj,Demmel:2015oqa,Ohta:2015fcu} or by evaluating the flow on maximally symmetric backgrounds with negative curvature, i.e. hyperbolic spaces \cite{Falls:2016msz}.

\subsection{$f(R,{\rm Ric}^2)$ Gravity}

We now turn to models which additionally allow for Ricci tensor interactions \cite{Falls:2017lst}, and consider gravitational 
Lagrangians of the form $\mathcal{L} =  f(R,{\rm Ric}^2)$, where
\begin{align}
\begin{split}
f(R,{\rm Ric}^2)&=F({\rm Ric}^2)+R \cdot Z({\rm Ric}^2)\,.
\end{split}
\end{align}
The functions $F$ and $Z$ are unspecified a priori, and characterise the even $\sim F(R)$ and odd $\sim R\cdot Z(R)$ parts of the action under reflection in field space $R\leftrightarrow -R$. In a polynomial expansion in the fields, the action contains the Einstein-Hilbert action to the lowest orders. The characteristic functions derived from this action are
\begin{equation}\label{RRic}
	\begin{split}
		L  &= F(x)+R \, Z(x) \, , \\
		E &=  \s0{1}{2} \left[ F' (x) + R\, Z'(x) \right] R+ Z (x) \, , \\
		A &=  \s0{1}{4} \left[ F'' (x) + R\, Z'' (x) \right] R^2  
		+  R\, Z' (x) \, ,\\
		B &= 2\,F' (x) + 2\,R\, Z'(x) \, , \\
		C &= 0\, ,
	\end{split}
\end{equation}
where $x=\s014 R^2$. Clearly, Ricci tensor interactions now contribute to the coefficients $L$ and $B$ while the coefficient $C$ remains trivial.
Within the asymptotic safety scenario, the functions  $F$ and $Z$ have been determined self-consistently by the requirement that an interacting fixed point is reached  in the ultraviolet \cite{Falls:2017lst}.

\subsection{$f(R, {\rm Riem}^2)$ Gravity}
In the same spirit, we consider gravitational actions which depend on Ricci scalar and Riemann tensor interactions \cite{Kluth:2020bdv}, 
but not on Ricci tensor ones, with a gravitational Lagrangian of the form $\mathcal{L} =  f(R,{\rm Riem}^2)$ where
\begin{align}\label{RRiem}
\begin{split}
f(R,{\rm Riem}^2)&=F({\rm Riem}^2)+R \cdot Z({\rm Riem}^2)\,.
\end{split}
\end{align}
Once more, the functions $F$ and $Z$ are unspecified a priori, and the action \eq{RRiem}  contains the Einstein-Hilbert action to the lowest orders in a polynomial expansion. The functions  $F$ and $Z$ have been determined self-consistently by the requirement that an interacting UV fixed point arises in the UV   \cite{Kluth:2020bdv}. The characteristic functions are found to be
\begin{equation}\label{ABC_RRiem}
	\begin{split}
		L  &= F(x)+R \, Z(x) \, , \\
		E &=  \s0{1}{3} \left[ F' (x) + R\, Z'(x) \right] R+ Z (x) \, , \\
		A &=  \s0{1}{9} \left[ F'' (x) + \s023 R\, Z'' (x) \right] R^2  
		+  R\, Z' (x) \, ,\\
		B &= 0\,,\\
		C &= 2\,F' (x) + 2\,R\, Z'(x) \,,
	\end{split}
\end{equation}
where $x=\s016 R^2$.  Notice that the absence of Ricci tensor interactions entails $B=0$. Using \eq{ABC_RRiem} together with the linear split $(\tau = 0)$ and specific endomorphism parameters \eq{e1e4} and \eq{e2e3}  
the flow equation \eq{floweqnabc} reduces to expressions given earlier in \cite{Kluth:2020bdv}.

\subsection{$f(R,{\rm Ric}^2, {\rm Riem}^2)$ Gravity}
\label{sec:paramquad}

The models of  the two preceeding sections can be combined by considering general Lagrangian of the form $\mathcal{L} = f(R,{\rm Ric}^2, {\rm Riem}^2)$ \cite{Falls:2017lst,Kluth:2020bdv}, where
\begin{align}
\begin{split}
f(R,{\rm Ric}^2, {\rm Riem}^2)&=F(\alpha \, R^2 +\beta\, {\rm Ric}^2+\gamma\, {\rm Riem}^2)+R \cdot Z(\alpha \, R^2 +\beta\, {\rm Ric}^2+\gamma\, {\rm Riem}^2)\,.
\end{split}
\end{align}
Besides the two free functions $F$ and $Z$, we have also introduced three free parameters $\alpha$, $\beta$ and $\gamma$ which characterise their argument. In practice, only two of the three parameters are independent, but for the derivation of expressions it is convenient to keep all three of them. The characteristic functions are then found to be
\begin{equation}\label{ABC_RRicRiem}
	\begin{split}
		L  &= F(x)+R \, Z(x) \, , \\
		E &=  \left(2 \alpha + \s0{1}{2}\beta + \s0{1}{3}\gamma \right) \left[ F' (x) + R\, Z'(x) \right] R+ Z (x) \, , \\
		A &=   2\, \alpha \left[ F'(x) + R \,Z' (x) \right] +
		 \left(2 \alpha + \s0{\beta }{2}+ \s0{\gamma}{3} \right)^2 \left[ F'' (x) + \left( 4 \alpha + \beta + \s0{2\gamma}{3} \right)  R\, Z'' (x) \right] R^2  
		+  R\, Z' (x)\\
		B &= 2\,\beta\left[F' (x) + R\, Z'(x) \right]\,,\\
		C &= 2\,\gamma\left[F' (x) + R\, Z'(x) \right]\,,
	\end{split}
\end{equation}
where $x=\left(\alpha+\s014\beta+\s016\gamma\right) R^2$. For this class of models, we note that the coefficients $B$ and $C$ are proportional to each other, $B/C=\beta/\gamma$, and non-zero in general, which permits  settings where spin-2 ghosts are absent from the outset $(B/C=-\s014)$, see \eq{eqn:ghostdof}. Using a linear split $(\tau = 0)$ and endomorphism parameters \eq{e1e4} and \eq{e2e3}  we reproduce the flow equation derived previously in \cite{Kluth:2020bdv}.

\subsection{Higher Order  Invariants}\label{sct:rezept}
Finally, we point out how our setup based on the action-independent form for the Hessians can be exploited for  investigations of quantum gravity, particularly clarifying the role of higher-order curvature invariants without necessarily starting from an explicit action.

Firstly, the flows of  gravitational actions which include different curvature invariants can be analysed by substituting appropriate values for the characteristic functions $B$, $C$, and $L$. Therefore, for each curvature invariant which does not contain covariant derivatives, it is possible to  identify the corresponding values for  the parameters $A,B,C,E$, and $L$ \cite{Bueno:2016ypa}. 
For convenience,  we tabulate in Tab.~\ref{tab:dim246operators} the  parameters 
for the first 38  curvature monomials not containing covariant derivatives, up to order four in curvature.
For notational convenience we express their values in terms of
\begin{equation}
	\Lambda = \frac{R}{d (d - 1)} \, ,
\end{equation}
and recall that parameters are functions of $R$. With these values at hand, a  practical recipe consists in studying the effects of the corresponding curvature invariants (or linear combinations thereof) by inserting the corresponding values (or  linear combinations thereof) into the  flow  \eq{floweqnabc}.

\begin{table*}[t]
	\centering
	\addtolength{\tabcolsep}{0pt}
	\setlength{\extrarowheight}{2pt}
\scalebox{0.75}{
	\rowcolors{1}{Gray}{white}
	\begin{tabular}{`c?ccccc`}
		\toprule
		\rowcolor{Yellow} \bf Curvature Invariants & $\bm{L}$ & $\bm{E}$ & $\bm{A}$ & $\bm{B}$ & $\bm{C}$ \\[1ex] \midrule
		$R$ & $(d-1) d \Lambda$ & $1$ & $0$ & $0$ & $0$ \\ \midrule
$R^2$ & $(d-1)^2 d^2 \Lambda ^2$ & $2 (d-1) d \Lambda$ & $2$ & $0$ & $0$ \\
$R{}^{\mu}{}^{\nu} R{}_{\mu}{}_{\nu}$ & $(d-1)^2 d \Lambda ^2$ & $2 (d-1) \Lambda$ & $0$ & $2$ & $0$ \\
$R{}^{\mu}{}^{\nu}{}^{\rho}{}^{\sigma} R{}_{\mu}{}_{\nu}{}_{\rho}{}_{\sigma}$ & $2 (d-1) d \Lambda ^2$ & $4 \Lambda$ & $0$ & $0$ & $2$ \\ \midrule
$R^3$ & $(d-1)^3 d^3 \Lambda ^3$ & $3 (d-1)^2 d^2 \Lambda ^2$ & $6 (d-1) d \Lambda$ & $0$ & $0$ \\
$R R{}^{\mu}{}^{\nu} R{}_{\mu}{}_{\nu}$ & $(d-1)^3 d^2 \Lambda ^3$ & $3 (d-1)^2 d \Lambda ^2$ & $4 (d-1) \Lambda$ & $2 (d-1) d \Lambda$ & $0$ \\
$R{}^{\nu}{}^{\rho} R{}_{\mu}{}_{\nu} R{}^{\mu}{}_{\rho}$ & $(d-1)^3 d \Lambda ^3$ & $3 (d-1)^2 \Lambda ^2$ & $0$ & $6 (d-1) \Lambda$ & $0$ \\
$R R{}^{\mu}{}^{\nu}{}^{\rho}{}^{\sigma} R{}_{\mu}{}_{\nu}{}_{\rho}{}_{\sigma}$ & $2 (d-1)^2 d^2 \Lambda ^3$ & $6 (d-1) d \Lambda ^2$ & $8 \Lambda$ & $0$ & $2 (d-1) d \Lambda$ \\
$R{}_{\rho}{}_{\sigma} R{}_{\mu}{}_{\nu} R{}^{\mu}{}^{\rho}{}^{\nu}{}^{\sigma}$ & $(d-1)^3 d \Lambda ^3$ & $3 (d-1)^2 \Lambda ^2$ & $2 \Lambda$ & $2 (2 d-3) \Lambda$ & $0$ \\
$R{}^{\nu}{}^{\alpha}{}^{\sigma}{}^{\beta} R{}_{\mu}{}_{\nu}{}_{\rho}{}_{\sigma} R{}^{\mu}{}_{\alpha}{}^{\rho}{}_{\beta}$ & $(d-2) (d-1) d \Lambda ^3$ & $3 (d-2) \Lambda ^2$ & $0$ & $6 \Lambda$ & $-3 \Lambda$ \\
$R{}^{\rho}{}^{\sigma}{}^{\alpha}{}^{\beta} R{}_{\mu}{}_{\nu}{}_{\rho}{}_{\sigma} R{}^{\mu}{}^{\nu}{}_{\alpha}{}_{\beta}$ & $4 (d-1) d \Lambda ^3$ & $12 \Lambda ^2$ & $0$ & $0$ & $12 \Lambda$ \\
$R{}^{\nu}{}^{\rho}{}^{\sigma}{}^{\alpha} R{}_{\mu}{}_{\nu} R{}^{\mu}{}_{\rho}{}_{\sigma}{}_{\alpha}$ & $2 (d-1)^2 d \Lambda ^3$ & $6 (d-1) \Lambda ^2$ & $0$ & $8 \Lambda$ & $2 (d-1) \Lambda$ \\ \midrule
$R^4$ & $(d-1)^4 d^4 \Lambda ^4$ & $4 (d-1)^3 d^3 \Lambda ^3$ & $12 (d-1)^2 d^2 \Lambda ^2$ & $0$ & $0$ \\
$R^2 R{}^{\mu}{}^{\nu} R{}_{\mu}{}_{\nu}$ & $(d-1)^4 d^3 \Lambda ^4$ & $4 (d-1)^3 d^2 \Lambda ^3$ & $10 (d-1)^2 d \Lambda ^2$ & $2 (d-1)^2 d^2 \Lambda ^2$ & $0$ \\
$R{}^{\rho}{}^{\sigma} R{}_{\rho}{}_{\sigma} R{}^{\mu}{}^{\nu} R{}_{\mu}{}_{\nu}$ & $(d-1)^4 d^2 \Lambda ^4$ & $4 (d-1)^3 d \Lambda ^3$ & $8 (d-1)^2 \Lambda ^2$ & $4 (d-1)^2 d \Lambda ^2$ & $0$ \\
$R R{}^{\nu}{}^{\rho} R{}_{\mu}{}_{\nu} R{}^{\mu}{}_{\rho}$ & $(d-1)^4 d^2 \Lambda ^4$ & $4 (d-1)^3 d \Lambda ^3$ & $6 (d-1)^2 \Lambda ^2$ & $6 (d-1)^2 d \Lambda ^2$ & $0$ \\
$R{}^{\nu}{}_{\sigma} R{}^{\rho}{}^{\sigma} R{}_{\mu}{}_{\nu} R{}^{\mu}{}_{\rho}$ & $(d-1)^4 d \Lambda ^4$ & $4 (d-1)^3 \Lambda ^3$ & $0$ & $12 (d-1)^2 \Lambda ^2$ & $0$ \\
$R^2 R{}^{\mu}{}^{\nu}{}^{\rho}{}^{\sigma} R{}_{\mu}{}_{\nu}{}_{\rho}{}_{\sigma}$ & $2 (d-1)^3 d^3 \Lambda ^4$ & $8 (d-1)^2 d^2 \Lambda ^3$ & $20 (d-1) d \Lambda ^2$ & $0$ & $2 (d-1)^2 d^2 \Lambda ^2$ \\
$R{}^{\rho}{}^{\sigma}{}^{\alpha}{}^{\beta} R{}_{\rho}{}_{\sigma}{}_{\alpha}{}_{\beta} R{}^{\mu}{}^{\nu} R{}_{\mu}{}_{\nu}$ & $2 (d-1)^3 d^2 \Lambda ^4$ & $8 (d-1)^2 d \Lambda ^3$ & $16 (d-1) \Lambda ^2$ & $4 (d-1) d \Lambda ^2$ & $2 (d-1)^2 d \Lambda ^2$ \\
$R{}^{\alpha}{}^{\beta}{}^{\gamma}{}^{\delta} R{}_{\alpha}{}_{\beta}{}_{\gamma}{}_{\delta} R{}^{\mu}{}^{\nu}{}^{\rho}{}^{\sigma} R{}_{\mu}{}_{\nu}{}_{\rho}{}_{\sigma}$ & $4 (d-1)^2 d^2 \Lambda ^4$ & $16 (d-1) d \Lambda ^3$ & $32 \Lambda ^2$ & $0$ & $8 (d-1) d \Lambda ^2$ \\
$R R{}_{\rho}{}_{\sigma} R{}_{\mu}{}_{\nu} R{}^{\mu}{}^{\rho}{}^{\nu}{}^{\sigma}$ & $(d-1)^4 d^2 \Lambda ^4$ & $4 (d-1)^3 d \Lambda ^3$ & $2 (d-1) (4 d-3) \Lambda ^2$ & $2 (d-1) d (2 d-3) \Lambda ^2$ & $0$ \\
$R{}_{\rho}{}_{\sigma} R{}^{\sigma}{}_{\alpha} R{}_{\mu}{}_{\nu} R{}^{\mu}{}^{\rho}{}^{\nu}{}^{\alpha}$ & $(d-1)^4 d \Lambda ^4$ & $4 (d-1)^3 \Lambda ^3$ & $4 (d-1) \Lambda ^2$ & $4 (d-1) (2 d-3) \Lambda ^2$ & $0$ \\
$R R{}^{\nu}{}^{\alpha}{}^{\sigma}{}^{\beta} R{}_{\mu}{}_{\nu}{}_{\rho}{}_{\sigma} R{}^{\mu}{}_{\alpha}{}^{\rho}{}_{\beta}$ & $(d-2) (d-1)^2 d^2 \Lambda ^4$ & $4 (d-2) (d-1) d \Lambda ^3$ & $6 (d-2) \Lambda ^2$ & $6 (d-1) d \Lambda ^2$ & $-3 (d-1) d \Lambda ^2$ \\
$R R{}^{\rho}{}^{\sigma}{}^{\alpha}{}^{\beta} R{}_{\mu}{}_{\nu}{}_{\rho}{}_{\sigma} R{}^{\mu}{}^{\nu}{}_{\alpha}{}_{\beta}$ & $4 (d-1)^2 d^2 \Lambda ^4$ & $16 (d-1) d \Lambda ^3$ & $24 \Lambda ^2$ & $0$ & $12 (d-1) d \Lambda ^2$ \\
$R R{}^{\nu}{}^{\rho}{}^{\sigma}{}^{\alpha} R{}_{\mu}{}_{\nu} R{}^{\mu}{}_{\rho}{}_{\sigma}{}_{\alpha}$ & $2 (d-1)^3 d^2 \Lambda ^4$ & $8 (d-1)^2 d \Lambda ^3$ & $12 (d-1) \Lambda ^2$ & $8 (d-1) d \Lambda ^2$ & $2 (d-1)^2 d \Lambda ^2$ \\
$R{}^{\nu}{}^{\alpha}{}^{\sigma}{}^{\beta} R{}_{\rho}{}_{\sigma} R{}_{\mu}{}_{\nu} R{}^{\mu}{}_{\alpha}{}^{\rho}{}_{\beta}$ & $2 (d-1)^3 d \Lambda ^4$ & $8 (d-1)^2 \Lambda ^3$ & $2 \Lambda ^2$ & $2 (9 d-10) \Lambda ^2$ & $2 (d-1)^2 \Lambda ^2$ \\
$R{}^{\nu}{}^{\sigma}{}^{\alpha}{}^{\beta} R{}_{\rho}{}_{\sigma} R{}_{\mu}{}_{\nu} R{}^{\mu}{}^{\rho}{}_{\alpha}{}_{\beta}$ & $2 (d-1)^3 d \Lambda ^4$ & $8 (d-1)^2 \Lambda ^3$ & $4 \Lambda ^2$ & $4 (4 d-5) \Lambda ^2$ & $2 (d-1)^2 \Lambda ^2$ \\
$R{}^{\rho}{}^{\alpha}{}_{\gamma}{}_{\delta} R{}^{\sigma}{}^{\beta}{}^{\gamma}{}^{\delta} R{}_{\mu}{}_{\nu}{}_{\rho}{}_{\sigma} R{}^{\mu}{}^{\nu}{}_{\alpha}{}_{\beta}$ & $4 (d-1) d \Lambda ^4$ & $16 \Lambda ^3$ & $0$ & $0$ & $24 \Lambda ^2$ \\
$R{}_{\rho}{}_{\sigma} R{}^{\rho}{}^{\alpha}{}^{\sigma}{}^{\beta} R{}_{\mu}{}_{\nu} R{}^{\mu}{}_{\alpha}{}^{\nu}{}_{\beta}$ & $(d-1)^4 d \Lambda ^4$ & $4 (d-1)^3 \Lambda ^3$ & $2 (3 d-4) \Lambda ^2$ & $2 (d (3 d-8)+6) \Lambda ^2$ & $0$ \\
$R{}^{\nu}{}_{\gamma}{}^{\alpha}{}_{\delta} R{}^{\sigma}{}^{\gamma}{}^{\beta}{}^{\delta} R{}_{\mu}{}_{\nu}{}_{\rho}{}_{\sigma} R{}^{\mu}{}_{\alpha}{}^{\rho}{}_{\beta}$ & $(d-1) d (3 d-5) \Lambda ^4$ & $4 (3 d-5) \Lambda ^3$ & $0$ & $28 \Lambda ^2$ & $(4 d-15) \Lambda ^2$ \\
$R{}^{\rho}{}_{\gamma}{}^{\alpha}{}_{\delta} R{}^{\sigma}{}^{\gamma}{}^{\beta}{}^{\delta} R{}_{\mu}{}_{\nu}{}_{\rho}{}_{\sigma} R{}^{\mu}{}^{\nu}{}_{\alpha}{}_{\beta}$ & $2 (d-2) (d-1) d \Lambda ^4$ & $8 (d-2) \Lambda ^3$ & $0$ & $20 \Lambda ^2$ & $2 (d-7) \Lambda ^2$ \\
$R{}^{\nu}{}_{\beta}{}^{\sigma}{}_{\gamma} R{}^{\rho}{}^{\beta}{}^{\alpha}{}^{\gamma} R{}_{\mu}{}_{\nu} R{}^{\mu}{}_{\rho}{}_{\sigma}{}_{\alpha}$ & $(d-2) (d-1)^2 d \Lambda ^4$ & $4 (d-2) (d-1) \Lambda ^3$ & $2 \Lambda ^2$ & $2 (5 d-9) \Lambda ^2$ & $-3 (d-1) \Lambda ^2$ \\
$R{}^{\rho}{}^{\sigma}{}_{\gamma}{}_{\delta} R{}^{\alpha}{}^{\beta}{}^{\gamma}{}^{\delta} R{}_{\mu}{}_{\nu}{}_{\rho}{}_{\sigma} R{}^{\mu}{}^{\nu}{}_{\alpha}{}_{\beta}$ & $8 (d-1) d \Lambda ^4$ & $32 \Lambda ^3$ & $0$ & $0$ & $48 \Lambda ^2$ \\
$R{}^{\nu}{}^{\rho}{}_{\beta}{}_{\gamma} R{}^{\sigma}{}^{\alpha}{}^{\beta}{}^{\gamma} R{}_{\mu}{}_{\nu} R{}^{\mu}{}_{\rho}{}_{\sigma}{}_{\alpha}$ & $4 (d-1)^2 d \Lambda ^4$ & $16 (d-1) \Lambda ^3$ & $0$ & $24 \Lambda ^2$ & $12 (d-1) \Lambda ^2$ \\
$R{}^{\nu}{}_{\sigma}{}_{\alpha}{}_{\beta} R{}^{\rho}{}^{\sigma}{}^{\alpha}{}^{\beta} R{}_{\mu}{}_{\nu} R{}^{\mu}{}_{\rho}$ & $2 (d-1)^3 d \Lambda ^4$ & $8 (d-1)^2 \Lambda ^3$ & $0$ & $20 (d-1) \Lambda ^2$ & $2 (d-1)^2 \Lambda ^2$ \\
$R{}^{\sigma}{}_{\beta}{}_{\gamma}{}_{\delta} R{}^{\alpha}{}^{\beta}{}^{\gamma}{}^{\delta} R{}_{\mu}{}_{\nu}{}_{\rho}{}_{\sigma} R{}^{\mu}{}^{\nu}{}^{\rho}{}_{\alpha}$ & $4 (d-1)^2 d \Lambda ^4$ & $16 (d-1) \Lambda ^3$ & $0$ & $32 \Lambda ^2$ & $8 (d-1) \Lambda ^2$ \\
$R{}^{\rho}{}_{\alpha}{}_{\beta}{}_{\gamma} R{}^{\sigma}{}^{\alpha}{}^{\beta}{}^{\gamma} R{}_{\mu}{}_{\nu} R{}^{\mu}{}_{\rho}{}^{\nu}{}_{\sigma}$ & $2 (d-1)^3 d \Lambda ^4$ & $8 (d-1)^2 \Lambda ^3$ & $8 \Lambda ^2$ & $4 (3 d-5) \Lambda ^2$ & $2 (d-1)^2 \Lambda ^2$ \\
$R{}^{\nu}{}_{\gamma}{}^{\sigma}{}_{\delta} R{}^{\alpha}{}^{\gamma}{}^{\beta}{}^{\delta} R{}_{\mu}{}_{\nu}{}_{\rho}{}_{\sigma} R{}^{\mu}{}_{\alpha}{}^{\rho}{}_{\beta}$ & $(d-1) d ((d-3) d+4) \Lambda ^4$ & $4 ((d-3) d+4) \Lambda ^3$ & $4 \Lambda ^2$ & $8 (d-3) \Lambda ^2$ & $10 \Lambda ^2$ \\
		\bottomrule
	\end{tabular}
}
	\caption{Shown are the parameters $L$, $E$, $A$, $B$, and $C$ corresponding to curvature invariants up to quartic order, not containing covariant derivatives.
	In the main text we mostly take $L$, $B$ and $C$ as the three independent parameters, with $E$ and $A$ determined through \eq{eqn:firstderivrel} and \eq{eqn:secderivrel}.}
	\label{tab:dim246operators}
\end{table*}

Secondly, one may also  start directly from the characteristic functions $L$, $B$, and $C$ without referring to any particular action $\mathcal{L} (\text{Riem})$ polynomial in curvature. To that end, consider a general Lagrangian of the form
\begin{equation}
	\mathcal{L} (\text{Riem}) = \sum_{n = 0}^\infty \blambda_n X_n \, ,
\end{equation}
containing arbitrary operators $X_n$ of order $n$ in curvature constructed from the Riemann tensor and the inverse metric. On a maximally symmetric background, any operator $X_n$ acquires the form
\begin{equation}
	X_n \big|_\text{msb} = L_n R^n \, ,
\end{equation}
with some spacetime dimension dependent constant $L_n$. Therefore, evaluating the Lagrangian on a maximally symmetric background, the corresponding scalar functions $L, B$ and $C$ have the following expansions in terms of the Ricci scalar curvature,
\begin{equation}\label{LBCexpansion}
\begin{aligned}
	L &= \sum_{n = 0}^\infty \blambda_n L_n R^n \,,\\
	B &= \sum_{n = 2}^\infty \blambda_n B_n R^{n - 2} \,, \\
	C &= \sum_{n = 2}^\infty \blambda_n C_n R^{n - 2} \,.
\end{aligned}
\end{equation}
Note that the sums for $B$ and $C$ start at $n = 2$, which can be understood following \eq{eqn:detABC}, in particular noticing that the Einstein-Hilbert terms cannot contribute to $B$ or $C$. 

So far, we have three functions $L$, $B$, and $C$ depending on four sets of parameters $\{ \blambda_n, L_n, B_n, C_n \}$, one of which is redundant. In fact, the numbers $L_n$ are redundant in that they correspond to the normalisation of operators and can always be absorbed into a rescaling of coupling constants.\footnote{This is possible for $L_n \neq 0$ which we require anyway following the arguments given in \sct{sct:projection}.} Therefore, we may introduce
\begin{equation}
\begin{aligned}
	\widetilde{\lambda}_n &= \blambda_n L_n \, , \qquad \\
	\widetilde{B}_n& = {B_n}/{L_n} \, , \qquad \\
	\widetilde{C}_n &= {C_n}/{L_n} \,,
\end{aligned}
	\label{eqn:characresc}
\end{equation}
such that
\begin{equation}
\begin{aligned}
	L  &= \sum_n \widetilde{\lambda}_n\, R^n \, , \qquad \\
	B &= \sum_{n = 2} \widetilde{\lambda}_n\, \widetilde{B}_n\, R^{n - 2} \, , \qquad \\
	C &= \sum_{n = 2} \widetilde{\lambda}_n\, \widetilde{C}_n\, R^{n - 2} \, .
\end{aligned}
	\label{eqn:Lfsbresc}
\end{equation}
Thus, on a maximally symmetric background we can map any action $\mathcal{L} (\text{Riem})$ to characteristic functions of the form \eq{eqn:Lfsbresc}. Consequently, we can study the effects of all possible higher curvature invariants (those which do not vanish on maximally symmetric backgrounds) by keeping the form of $L (R)$ fixed according to \eq{eqn:Lfsbresc}, while varying  the parameters $\{\widetilde{B}_n, \widetilde{C}_n \}$. In general, these parameters can take arbitrary  values along the real axis, and in particular they are not bounded.

As an example for this idea, consider the class of actions introduced in \sct{sec:paramquad} at quadratic level in curvature. In this case, the Lagrangian takes the form
\begin{equation}
	\mathcal{L} (\text{Riem}) = \blambda_0 + \blambda_1 R + \blambda_2 \left( \alpha R^2 + \beta R_{\mu \nu} R^{\mu \nu} + \gamma R_{\rho \sigma \mu \nu} R^{\rho \sigma \mu \nu} \right) \,.
\end{equation} 
On a maximally symmetric background we find
\begin{equation}
\begin{aligned}
	L &= \, \blambda_0 + \blambda_1 R + \blambda_2 R^2 \left( \alpha + \s014 \beta + \s016 \gamma\right) \, , \qquad \\
	B &= \, 2 \beta \blambda_2 \, , \qquad\\
	 C &= \, 2 \gamma \blambda_2 \,.
\end{aligned}
\end{equation}
Apart from coupling constants, this model depends on three parameters $\alpha$, $\beta$, and $\gamma$. Following \eq{LBCexpansion}, \eq{eqn:characresc} and \eq{eqn:Lfsbresc}, an overall normalisation factor can be rescaled into $\blambda_2$ without changing the physical content of the model. Taking
\begin{equation}
\begin{aligned}
	\widetilde{\lambda}_2& = \left( \alpha + \s014 \beta + \s016 \gamma \right) \lambda_2 \, , \qquad \\
	b_2&= {\beta}/({\alpha + \s014 \beta + \s016 \gamma}) \, , \qquad \\
	c_2 &={\gamma}/({\alpha + \s014 \beta + \s016 \gamma})\, ,
	\label{eqn:repexample}
\end{aligned}
\end{equation}
leads to
\begin{equation}
\begin{aligned}
	L &= \, \lambda_0 + \lambda_1\, R + \widetilde{\lambda}_2\, R^2 \, , \qquad \\
	B &= \, 2\, b_2\,
	\widetilde{\lambda}_2 \, , \qquad \\
	C &= \, 2\, c_2
	\, \widetilde{\lambda}_2 \, .
\end{aligned}
\end{equation}
The rescaling \eq{eqn:repexample} has eliminated one parameter from the three-parameter family of actions we started with, and we have ended up  with a two-parameter family of RG flows with  $(b_2,c_2)$ characterising general fourth-order flows with quantum fluctuations evaluated on  spheres. 

This idea  can naturally be carried over for actions containing arbitrary higher curvature invariants \eq{eqn:Lfsbresc}, leaving us with  at most two free parameters $\{\widetilde{B}_n, \widetilde{C}_n \}$ for every order  $n\ge 2$ in curvature monomials. It will be interesting to apply these setups for systematic fixed point searches in higher order theories of gravity, which is left for future work.

\section{\bf Discussion and Outlook}\label{sec:conclusions}

We have put forward new functional renormalisation group flows for  $f({R}_{\mu \nu\rho \sigma })$ quantum gravity, in any dimension. The most important novelty is that the underlying Lagrangian for these types of theories can be taken to be  {\it any} function of the  Riemann tensor and the inverse metric.
As such, our setup offers a change of perspective in that functional flows can now be determined without the need to specify the underlying Langrangian  beyond the particular form $\sim f({R}_{\mu \nu\rho \sigma })$. 

To achieve the result, crucially, full advantage has been taken of maximally symmetric backgrounds, conveniently employed  for the evaluation of operator traces. 
In consequence, the functional flows   \eq{eqn:flowdimful}, \eq{floweqnabc} are   characterised by  three independent  scalar functions, \eq{ABCdef}, which we have taken to be
 the Lagrangian evaluated on the background, $L$, and  two quantities $B$ and $C$, which, respectively, account for  effects due to Ricci and Riemann tensor fluctuations. 
 On the technical side, we mostly followed standard choices  in the literature to achieve the explicit flow \eq{eqn:flowdimful}, \eq{floweqnabc}, 
 We also implemented  an interpolation between the popular linear and exponential metric splits  \eq{eqn:taumetricsplit}. 
 Our setup is highly flexible  and allows the full range of choices for e.g.~cutoff types and  shape functions, gauge fixings, endomorphism parameters, and more, and all of this in combination with  heat kernel expansions \cite{Kluth:2019vkg} or spectral sum techniques.

 Overall, the new flow equation encompasses all   models on maximally symmetric backgrounds investigated previously within the asymptotic safety programme, to which it reduces for the corresponding parameter choices.
 What's more, the generality and structure of the setup  opens up a wide range of new applicabilities. 
 First and foremost,  it allows the study of quantum gravitational effects in  a plethora of new extensions beyond  Einstein gravity, polynomial or otherwise,  many of which  have hitherto been out of reach. Further, it enables qualitatively new types of fixed point search strategies within the operator space spanned by polynomial curvature invariants (Tab.~\ref{tab:dim246operators}), including horizontal (or vertical) searches across  curvature invariants with the same (or different) canonical mass dimensions.  Finally, we emphasise that the setup permits the study of quantum effects in extensions of general relativity relevant for cosmology and the physics of black holes. 
 We thus look forward to detailed explorations of the  landscape for asymptotically safe $f({R}_{\mu \nu\rho \sigma })$ theories.\\[4ex]

\centerline{\bf Acknowledgments}
\noindent
This work is supported by the Science Technology and Facilities Council (STFC) under the Studentship Grant  ST/S505766/1 (YK) 
and  the Consolidated Grant ST/T00102X/1 (DL).
\onecolumngrid

\setcounter{section}{0}

\section*{\bf Appendices}
\subsection{Metric Derivatives}
\label{sct:metricderiv}
In this Appendix, we  take care of the metric derivatives required for the evaluation of  \eq{eqn:firstsecvareq}. Following a line of reasoning put forward in \cite{Bueno:2016ypa,Padmanabhan:2011ex}, we start by considering an infinitesimal coordinate transformation
$x_\mu \rightarrow x_\mu + \xi_\mu (x)$.
Since the Lagrangian is a scalar, the variation under this coordinate transformation can be written as a Lie derivative
\begin{equation}
	\begin{split}
		\delta \mathcal{L} &= \xi^\eta \nabla_\eta \mathcal{L} (\text{Riem}) = \xi^\eta \left( \mathcal{W}^{\rho \sigma \mu \nu} \nabla_\eta R_{\rho \sigma \mu \nu} + \pderiv{\mathcal{L} (\text{Riem})}{g^{\mu \nu}} \nabla_\eta g^{\mu \nu} \right) \\
		&= \xi^\eta \mathcal{W}^{\rho \sigma \mu \nu} \nabla_\eta R_{\rho \sigma \mu \nu} \, .
	\end{split}
	\label{eqn:padfirst}
\end{equation}
where we recall that
\begin{equation}
	\mathcal{W}^{\rho \sigma \mu \nu} \equiv \pderiv{\mathcal{L} (\text{Riem})}{R_{\rho \sigma \mu \nu}} \,.
\end{equation}
On the other hand, the change in $\mathcal{L}$ can also be expressed in terms of the changes in the Riemann tensor and the metric,
\begin{equation}
	\delta \mathcal{L} = \mathcal{W}^{\rho \sigma \mu \nu} \delta R_{\rho \sigma \mu \nu} + \pderiv{\mathcal{L} (\text{Riem})}{g^{\mu \nu}} \delta g^{\mu \nu} \, .
	\label{eqn:padsec1}
\end{equation}
With the change of the inverse metric and the Riemann tensor  given by
\begin{align}
	\delta g^{\mu \nu} &= -\nabla^\mu \xi^\nu - \nabla^\nu \xi^\mu \,,\\
		\delta R_{\rho \sigma \mu \nu} &= \xi^\eta \nabla_\eta R_{\rho \sigma \mu \nu} + (\nabla_\rho \xi^\eta) R_{\eta \sigma \mu \nu} + (\nabla_\sigma \xi^\eta) R_{\rho \eta \mu \nu} + (\nabla_\mu \xi^\eta) R_{\rho \sigma \eta \nu} + (\nabla_\nu \xi^\eta) R_{\rho \sigma \mu \eta} \, ,
\end{align}
and also using the symmetries of $\mathcal{W}^{\rho \sigma \mu \nu}$ we can recast \eq{eqn:padsec1} into the form
\begin{equation}
	\delta \mathcal{L} = \mathcal{W}^{\rho \sigma \mu \nu} (\xi^\eta \nabla_\eta R_{\rho \sigma \mu \nu} + 4 (\nabla_\rho \xi^\eta) R_{\eta \sigma \mu \nu}) - 2 \pderiv{\mathcal{L} (\text{Riem})}{g^{\mu \nu}} \nabla^\mu \xi^\nu \, .
	\label{eqn:padsec}
\end{equation}
Equating \eq{eqn:padfirst} with \eq{eqn:padsec} we arrive at
\begin{equation}
	0 =  
	\left[2\mathcal{W}_\rho^{\ \sigma \mu \nu} R_{\eta \sigma \mu \nu} - \pderiv{\mathcal{L} (\text{Riem})}{g^{\rho \eta}}\right]  \nabla^\rho \xi^\eta \,.
\end{equation}
As this  must hold true for any $\xi$, we conclude that the first derivative of the Lagrangian with respect to the metric, and written  in terms of $\mathcal{W}^{\rho \sigma \mu \nu}$, is given by 
\begin{equation}
	\pderiv{\mathcal{L} (\text{Riem})}{g^{\lambda \eta}} = 2 g_{\rho ( \lambda} R_{\eta) \sigma \mu \nu} \mathcal{W}^{\rho \sigma \mu \nu} \,.
\end{equation}
While the derivation of \eq{eqn:LgdR} has made use of the Lagrangian $\mathcal{L}$ being solely a function of the Riemann tensor and the metric field,  
we have not made any choice for the background metric. Therefore, the result \eq{eqn:LgdR} is valid for 
general geometries.

To obtain higher derivatives, we first take a derivative of \eq{eqn:LgdR} with respect to the Riemann tensor,
\begin{equation}
	\begin{split}
		\pderiv{^2\mathcal{L}(\text{Riem})}{R_{\rho \sigma \mu \nu}\partial  g^{\alpha \beta}} =& \, 2 \mathcal{W}_{( \alpha}^{\ \kappa \eta \xi} \mathcal{C}_{\beta) \kappa \eta \xi}^{\ \ \ \ \ \ \rho \sigma \mu \nu} + 2 g_{\zeta (\alpha} R_{\beta) \kappa \eta \xi} \pderiv{\mathcal{W}^{\zeta \kappa \eta \xi}}{R_{\rho \sigma \mu \nu}} \\
		=& \, g_{(\beta}^{\ \ [ \rho} \mathcal{W}_{\alpha)}^{\ \ \sigma ] \mu \nu} + g_{(\beta}^{\ \ [ \mu} \mathcal{W}_{\alpha)}^{\ \ \nu ] \rho \sigma} + 2 g_{\zeta (\alpha} R_{\beta) \kappa \eta \xi} \pderiv{\mathcal{W}^{\zeta \kappa \eta \xi}}{R_{\rho \sigma \mu \nu}} \,.
	\end{split}
\end{equation}
Here, we used
\begin{equation}
	\pderiv{R_{\rho \sigma \mu \nu}}{R_{\alpha \beta \gamma \delta}} = \mathcal{C}_{\rho \sigma \mu \nu}^{\ \ \ \ \ \alpha \beta \gamma \delta} \, ,
\end{equation}
with $\mathcal{C}_{\rho \sigma \mu \nu}^{\ \ \ \ \ \alpha \beta \gamma \delta}$ defined in \eq{Cdef}.
Taking a further metric derivative of \eq{eqn:LgdR} gives
\begin{equation}
	\pderiv{^2 \mathcal{L} (\text{Riem})}{g^{\rho \sigma} \partial g^{\mu \nu}} = - 2 g_{\alpha (\mu} g_{\nu) (\rho} \mathcal{W}^{\alpha \beta \gamma \delta} R_{\sigma) \beta \gamma \delta} + 2 \pderiv{\mathcal{W}^{\alpha \beta \gamma \delta}}{g^{\mu \nu}} g_{\alpha (\rho} R_{\sigma) \beta \gamma \delta} \, .
\end{equation}
This concludes the derivation of \eq{eqn:LgdR}, \eq{eqn:LgdRdR}, and \eq{eqn:LgdRdRdg} given in the main text.
It allows us to eliminate all derivatives with respect to the metric in \eq{eqn:firstsecvareq} in favour of Riemann derivatives. The latter can be parametrised in terms of four scalar functions on maximally symmetric backgrounds as seen in \sct{sct:riemderiv}.

\subsection{Hessians without York Decomposition}
\label{app:Hessian}

Here we present the Hessian of an of the form \eq{eqn:bGammaform} 
without making use of the York decomposition. Using the metric split \eq{eqn:taumetricsplit} we arrive at
\begin{equation}
	\begin{split}
		\left. \delta^2 \left( \bGamma_k \right) \right|_{\rm msb} =& \, \int {\rm d}^d x \sqrt{g} \, \Bigg\{ h \bigg[ \frac{R^2}{d^2} \left( \frac{d^2-4 d+2}{d (d-1)^2} B - \frac{1}{d} C + L'' \right) - \frac{d-2}{d-1} \frac{R}{d} L' + \frac{1}{4} L \\
		& \quad + \left( \frac{R}{d} \left( \frac{d^2+3 d-16}{4d(d-1)} B + \frac{2}{d(d-1)} C + 2 L'' \right) - \frac{1}{2} L' \right) \nabla^2 \\
		& \quad+ \left( \frac{d^2-d-8}{4d(d-1)} B - \frac{1}{d} C + L'' \right) \nabla^4 \bigg] h \\
		& + h_{\mu \nu} \bigg[ \frac{R^2}{(d - 1) d^2} \left( \frac{1}{d-1} B + 2 C \right) - \frac{R}{d (d - 1)} L' + (\tau -1) \left( \frac{L}{2} - \frac{R}{d} L' \right) \\
		& \quad + \left( - \frac{R}{d(d - 1)} \bigg( B + (d+1) C \bigg)+ \frac{1}{2} L' \right) \nabla^2 + \left( \frac{B}{4} + C \right) \nabla^4 \bigg] h^{\mu \nu} \\
		& + \left[ \frac{R}{d} \left( \frac{1}{2} B + \frac{4}{d-1} C \right) + L' \right] \left( \nabla_\mu h^{\mu \rho} \right) \left( \nabla^\nu h_{\nu \rho} \right) \\
		& + \left[ \frac{R}{d} \left( \frac{4}{(d-1) d} B + \frac{2 (3 d-1)}{(d-1) d} C - 2 L'' \right) + L' \right] h \left( \nabla_\mu \nabla_\nu h^{\mu \nu} \right) \\
		& + \left[ \frac{\left(d^2-d-4\right)}{2 (d-1) d} B + \frac{(d-1)}{d} C + L'' \right] \left( \nabla_\mu \nabla_\nu h^{\mu \nu} \right) \left( \nabla_\rho \nabla_\sigma h^{\rho \sigma} \right) \\
		& + \left[ \frac{B}{2} + 2 C \right] \left( \nabla^\mu h_{\mu \rho} \right) \left( \nabla^2 \nabla_\nu h^{\nu \rho} \right)  \\
		& + \left[ -\frac{\left(d^2-d-8\right)}{2 (d-1) d} B + \frac{2}{d} C - 2 L'' \right] h \nabla^2 \nabla_\mu \nabla_\nu h^{\mu \nu} \Bigg\} \, .
	\end{split}
	\label{eqn:genhess}
\end{equation}

\subsection{Fluctuation Integrals}
\label{sec:floweq}
Here, we give the coefficient functions $P$ and $D$ for the flow equation derived in \sct{sct:floweqmain}. In dimensionless quantities the flow equation is given by
\begin{equation}
	 \partial_t \ell +d \, \ell - 2 r \ell' = I [\ell, b, c] \,,
\end{equation}
with
\begin{equation}
	\begin{split}
		I [\ell, b, c] =& \, I_0 [\ell, b, c] + I_1 [\ell, b, c] \partial_t \ell' + I_2 [\ell, b, c] \partial_t \ell''+ I_3 [\ell, b, c] \partial_t b + I_4 [\ell, b, c] \partial_t c \, ,
	\end{split}
\end{equation}
Since we are using the York decomposition throughout the whole computation, we keep track of terms originating from the different York modes. This allows us to write the $I_i [\ell, b, c]$ as
\begin{subequations}
\begin{align}\label{In}
		I_0 [\ell, b, c] =& \frac{1}{\kappa_d} \left[ \frac{P_0^T [\ell, b, c]}{D^T [\ell, b, c]} + \frac{P_0^S [\ell, b, c]}{D^S [\ell, b, c]} - \frac{P^V_c}{D^V_c} - \frac{P^S_c}{D^S_c} \right] \, , \\
		I_1 [\ell, b, c] =& \frac{1}{\kappa_d} \left[ \frac{P_1^T}{D^T [\ell, b, c]} + \frac{P_1^S}{D^S [\ell, b, c]} \right] \, , \\
		I_2 [\ell, b, c] =& \frac{1}{\kappa_d} \frac{P_2^S}{D^S [\ell, b, c]} \, , \\
		I_3 [\ell, b, c] =& \frac{1}{\kappa_d} \left[ \frac{P_3^T}{D^T [\ell, b, c]} + \frac{P_3^S}{D^S [\ell, b, c]} \right] \, , \qquad \\
		I_4 [\ell, b, c] =& \frac{1}{\kappa_d} \left[ \frac{P_4^T}{D^T [\ell, b, c]} + \frac{P_4^S}{D^S [\ell, b, c]} \right] \, ,
\end{align}
\end{subequations}
where $T$, $V$ and $S$ denote contributions from tensorial, vectorial and scalar origin, respectively. The subscript $c$ refers to contributions from ghosts and Jacobians and
\begin{equation}
	\kappa_d = R^{-d/2} \,{\text{Vol}_d} \xrightarrow{d = 4} 384 \pi^2 \, ,
\end{equation}
originates from the volume integral on the left-hand side of \eq{eqn:wetterichbga}. 
The denominators $D$ are directly related to the denominators in \eq{fRG}, while the numerators $P$ originate from numerators in \eq{fRG} which have been split up to collect renormalisation scale derivatives according to \eq{eqn:flowres}.
The explicit coefficients $P$ and $D$ are rather lengthy, and we only give them for four dimensional spacetime, $d= 4$.

The denominators in \eq{In} are defined as
\begin{subequations}
	\begin{align}\label{DT}
		\begin{split}
			D^T [\ell, b, c] =& \, \left( \tau -1 \right) \ell - \left( -e_1 r+\frac{\tau}{2}  r-\frac{r}{3}+1 \right) \ell' + \left( -\frac{e_1 r^2}{6}+\frac{e_1^2 r^2}{2}-e_1 r+\frac{r^2}{72}+\frac{r}{6}+\frac{1}{2} \right) b \\
			& + \left(-\frac{5 e_1 r^2}{6}+2 e_1^2 r^2-4 e_1 r+\frac{r^2}{12}+\frac{5 r}{6}+2\right) c \, ,
		\end{split} \\
		\begin{split}\label{DS}
			D^S [\ell, b, c] =& \, \left( \tau + 1 \right) \ell - \left( \frac{3 e_4 r}{2}+\frac{\tau}{2}  r+r-\frac{3}{2} \right) \ell' + \left( 3 e_4 r^2+\frac{9 e_4^2 r^2}{2}-9 e_4 r+\frac{r^2}{2}-3 r+\frac{9}{2} \right) \ell'' \\
			& + \left(\frac{e_4 r^2}{4}+\frac{3 e_4^2 r^2}{4}-\frac{3 e_4 r}{2}-\frac{r}{4}+\frac{3}{4}\right) \left( \frac{b}{2} + c \right) \,.
		\end{split}
	\end{align}
\end{subequations}
The numerators  $P_0$ appearing  in \eq{In}  can be written as
\begin{subequations}
	\begin{align}
	\label{PT}
		P_0^T [\ell, b, c] &= P_0^{T \ell1} \ell' + P_0^{T \ell2} \ell'' + P_0^{T b0} b + P_0^{T b1} b' + P_0^{T c0} c + P_0^{T c1} c' \, , \\
		\label{PS}
		P_0^S [\ell, b, c] &= P_0^{S \ell1} \ell' + P_0^{S \ell2} \ell'' + P_0^{S \ell3} \ell''' + P_0^{S 0} \left( \tfrac{b}{2} + c \right) + P_0^{S 1} \left( \tfrac{b'}{2} + c' \right) \, ,
	\end{align}
\end{subequations}
with coefficient functions
\begin{subequations}
	\label{floweqnI0}
	\begin{align}
		P_0^{T \ell1} =& \, -\frac{e_1 r^3}{18}+10 e_1^2 r^3-40 e_1 r^2+20 e_1^3 r^3-120 e_1^2 r^2+180 e_1 r-\frac{311 r^3}{2268}+\frac{r^2}{9}+30 r-80 \, , \\
		P_0^{T \ell2} =& \, \frac{e_1 r^4}{18}-10 e_1^2 r^4+20 e_1 r^3-20 e_1^3 r^4+60 e_1^2 r^3-60 e_1 r^2+\frac{311 r^4}{2268}-\frac{r^3}{18}-10 r^2+20 r \, , \\
		P_0^{T b0} =& \, \frac{61 e_1 r^3}{18}-10 e_1^2 r^3+20 e_1 r^2-60 e_1^3 r^3+180 e_1^2 r^2-180 e_1 r-\frac{r^3}{108}-\frac{61 r^2}{18}-10 r+60 \, , \\
		\begin{split}
			P_0^{T b1} =& \, -\frac{e_1 r^5}{108}+\frac{61}{36} e_1^2 r^5-\frac{61 e_1 r^4}{18}-\frac{10}{3} e_1^3 r^5+10 e_1^2 r^4-10 e_1 r^3-15 e_1^4 r^5+60 e_1^3 r^4 \\
			& -90 e_1^2 r^3+60 e_1 r^2-\frac{1135 r^5}{54432}+\frac{r^4}{108}+\frac{61 r^3}{36}+\frac{10 r^2}{3}-15 r  \, ,
			\label{eqn:P0Tb1}
		\end{split} \\
		\begin{split}
			P_0^{T c0} =& \, \frac{152 e_1 r^3}{9}-30 e_1^2 r^3+60 e_1 r^2-240 e_1^3 r^3+720 e_1^2 r^2-720 e_1 r-\frac{5 r^3}{108}-\frac{152 r^2}{9}-30 r \\
			& +240 \, , 
		\end{split} \\
		\begin{split}
			P_0^{T c1} =& \, -\frac{5 e_1 r^5}{108}+\frac{76}{9} e_1^2 r^5-\frac{152 e_1 r^4}{9}-10 e_1^3 r^5+30 e_1^2 r^4-30 e_1 r^3-60 e_1^4 r^5+240 e_1^3 r^4 \\
			& -360 e_1^2 r^3+240 e_1 r^2-\frac{241 r^5}{2268}+\frac{5 r^4}{108}+\frac{76 r^3}{9}+10 r^2-60 r \, ,
		\end{split} \\
		P_0^{S \ell1} =& \, -\frac{29 e_4 r^3}{60}+3 e_4^2 r^3-12 e_4 r^2-6 e_4^3 r^3+36 e_4^2 r^2-54 e_4 r+\frac{37 r^3}{1512}+\frac{29 r^2}{30}+9 r+24 \, , \\
		\begin{split}
			P_0^{S \ell2} =& \, \frac{29 e_4 r^4}{60}-3 e_4^2 r^4+\frac{151 e_4 r^3}{10}+6 e_4^3 r^4-18 e_4^2 r^3+18 e_4 r^2-108 e_4^3 r^3+324 e_4^2 r^2\\
			& -324 e_4 r-\frac{37 r^4}{1512}-\frac{29 r^3}{20}-\frac{121 r^2}{10}-6 r+108 \, ,
		\end{split} \\
		\begin{split}
			P_0^{S \ell3} =& \, -\frac{29 e_4 r^5}{30}+\frac{91}{20} e_4^2 r^5-\frac{91 e_4 r^4}{10}-27 e_4^4 r^5+108 e_4^3 r^4-162 e_4^2 r^3+108 e_4 r^2+\frac{181 r^5}{3360} \\
			& +\frac{29 r^4}{30}+\frac{91 r^3}{20}-27 r \, , 
			\label{lppprefac}
		\end{split} \\
		P_0^{S 0} =& \, \frac{31 e_4 r^3}{60}+3 e_4^2 r^3-6 e_4 r^2-18 e_4^3 r^3+54 e_4^2 r^2-54 e_4 r-\frac{29 r^3}{360}-\frac{31 r^2}{60}+3 r+18 \, , \\
		\begin{split}
			P_0^{S 1} =& \, -\frac{29 e_4 r^5}{360}+\frac{31}{120} e_4^2 r^5-\frac{31 e_4 r^4}{60}+e_4^3 r^5-3 e_4^2 r^4+3 e_4 r^3-\frac{9 e_4^4 r^5}{2}+18 e_4^3 r^4-27 e_4^2 r^3\\
			& +18 e_4 r^2+\frac{127 r^5}{25920}+\frac{29 r^4}{360}+\frac{31 r^3}{120}-r^2-\frac{9 r}{2} \, .
		\end{split}
	\end{align}
\end{subequations}
The  numerators  $P_{1,2,3,4}$ appearing  in \eq{In}  take the form
 \begin{subequations}\label{P1}
	\begin{align}
		P_1^T =& \, -\frac{e_1 r^3}{36}+5 e_1^2 r^3-10 e_1 r^2+10 e_1^3 r^3-30 e_1^2 r^2+30 e_1 r-\frac{311 r^3}{4536}+\frac{r^2}{36}+5 r-10 \, , \\
		P_1^S =& \, -\frac{29 e_4 r^3}{120}+\frac{3 e_4^2 r^3}{2}-3 e_4 r^2-3 e_4^3 r^3+9 e_4^2 r^2-9 e_4 r+\frac{37 r^3}{3024}+\frac{29 r^2}{120}+\frac{3 r}{2}+3 \,,
		\end{align}
\end{subequations}
\begin{subequations}\label{P2}
	\begin{align}
		\begin{split}
			P_2^S =& \, \frac{29 e_4 r^4}{60}-\frac{91}{40} e_4^2 r^4+\frac{91 e_4 r^3}{20}+\frac{27 e_4^4 r^4}{2}-54 e_4^3 r^3+81 e_4^2 r^2-54 e_4 r-\frac{181 r^4}{6720}-\frac{29 r^3}{60} \\
			& -\frac{91 r^2}{40}+\frac{27}{2}\,,
		\end{split} \\
		\begin{split}
			P_3^T =& \, \frac{e_1 r^4}{216}-\frac{61}{72} e_1^2 r^4+\frac{61 e_1 r^3}{36}+\frac{5 e_1^3 r^4}{3}-5 e_1^2 r^3+5 e_1 r^2+\frac{15 e_1^4 r^4}{2}-30 e_1^3 r^3+45 e_1^2 r^2 \\
			& -30 e_1 r+\frac{1135 r^4}{108864}-\frac{r^3}{216}-\frac{61 r^2}{72}-\frac{5 r}{3}+\frac{15}{2} \, , 
		\end{split} \\
		\begin{split}
			P_3^S =& \, \frac{29 e_4 r^4}{1440}-\frac{31}{480} e_4^2 r^4+\frac{31 e_4 r^3}{240}-\frac{e_4^3 r^4}{4}+\frac{3 e_4^2 r^3}{4}-\frac{3 e_4 r^2}{4}+\frac{9 e_4^4 r^4}{8}-\frac{9 e_4^3 r^3}{2}+\frac{27 e_4^2 r^2}{4}\\
			& -\frac{9 e_4 r}{2}-\frac{127 r^4}{103680}-\frac{29 r^3}{1440}-\frac{31 r^2}{480}+\frac{r}{4}+\frac{9}{8} \, .
		\end{split} \\
		\begin{split}
			P_4^T =& \, \frac{5 e_1 r^4}{216}-\frac{38}{9} e_1^2 r^4+\frac{76 e_1 r^3}{9}+5 e_1^3 r^4-15 e_1^2 r^3+15 e_1 r^2+30 e_1^4 r^4-120 e_1^3 r^3+180 e_1^2 r^2 \\
			& -120 e_1 r+\frac{241 r^4}{4536}-\frac{5 r^3}{216}-\frac{38 r^2}{9}-5 r+30 \, , 
		\end{split} \\
		\begin{split}
			P_4^S =& \, \frac{29 e_4 r^4}{720}-\frac{31}{240} e_4^2 r^4+\frac{31 e_4 r^3}{120}-\frac{e_4^3 r^4}{2}+\frac{3 e_4^2 r^3}{2}-\frac{3 e_4 r^2}{2}+\frac{9 e_4^4 r^4}{4}-9 e_4^3 r^3+\frac{27 e_4^2 r^2}{2}\\
			& -9 e_4 r-\frac{127 r^4}{51840}-\frac{29 r^3}{720}-\frac{31 r^2}{240}+\frac{r}{2}+\frac{9}{4} \,,
		\end{split}
	\end{align}
\end{subequations}

Finally, we give the universal contributions coming from the auxiliary part. The denominators are given by
\begin{subequations}
	\begin{align}
		D^V_c = 1-e_2 r-\frac{r}{4} \, , \\
		D^S_c = 1-e_3 r-\frac{r}{3} \, ,
	\end{align}
	\label{eqn:techpolesdenom}
\end{subequations}
and the numerators by
\begin{subequations}
	\begin{align}
		P^V_c = 6 e_2 r^2-36 e_2^2 r^2+72 e_2 r+\frac{607 r^2}{60}-6 r-36 \, , \\
		P^S_c = 4 e_3 r^2-12 e_3^2 r^2+24 e_3 r+\frac{511 r^2}{90}-4 r-12 \, .
	\end{align}
\end{subequations}

\bibliography{bibtex,bib-Litim-Jan2022}
\bibliographystyle{mystyle}

\end{document}

%% file: layout.tex
\renewcommand{\baselinestretch}{1.}
\AtBeginDocument{
\heavyrulewidth=.16em
\lightrulewidth=.1em
\cmidrulewidth=.03em
\belowrulesep=.4ex
\belowbottomsep=0pt
\aboverulesep=.4ex
\abovetopsep=0pt
\cmidrulesep=\doublerulesep
\cmidrulekern=.5em
\defaultaddspace=.5em
}

\renewcommand{\baselinestretch}{1.1}
\long\def\del #1 \enddel { }

\renewcommand{\baselinestretch}{1.}

\makeatletter
    \def\CT@@do@color{%
      \global\let\CT@do@color\relax
            \@tempdima\wd\z@
            \advance\@tempdima\@tempdimb
            \advance\@tempdima\@tempdimc
    \advance\@tempdimb\tabcolsep
    \advance\@tempdimc\tabcolsep
    \advance\@tempdima2\tabcolsep
            \kern-\@tempdimb
            \leaders\vrule
                    \hskip\@tempdima\@plus  1fill
            \kern-\@tempdimc
            \hskip-\wd\z@ \@plus -1fill }

\definecolor{Gray}{gray}{0.85}
\definecolor{LightGray}{gray}{0.93}
\definecolor{LightGreen}{rgb}{0.88, 1, 0.88}
\definecolor{LightCyan}{rgb}{0.88,1,1}
\definecolor{LightRed}{rgb}{1, 0.85, 0.85}
\definecolor{LightYellow}{rgb}{1, 1, 0.85}
\definecolor{Yellow}{rgb}{1,1,0.05}
\definecolor{LightBlue}{rgb}{0.87, 0.94, 1}
\definecolor{white}{gray}{1}

\newcolumntype{G}{>{\columncolor{LightGray}}c}
\newcolumntype{C}{>{$}c<{$}}
\newcolumntype{?}{!{\vrule width 1pt}}
\newcolumntype{`}{!{\vrule width 1.5pt}}

%% file: commands.tex
\newcommand{\eq}[1]{(\ref{#1})}

\newcommand{\fig}[1]{Fig.\,\ref{#1}}

\newcommand{\sct}[1]{Sect.\,\ref{#1}}

\newcommand{\pderiv}[2]{\frac{\partial #1}{\partial #2}}

\newcommand{\fderiv}[2]{\frac{\delta #1}{\delta #2}}

\newcounter{notecount}

\newcommand{\beq}{\begin{equation}}
\newcommand{\eeq}{\end{equation}}
\newcommand{\bsp}{\begin{split}}
\newcommand{\esp}{\end{split}}
\newcommand{\bfi}{\begin{figure}}
\newcommand{\efi}{\end{figure}}
\def\bea{\arraycolsep .1em \begin{eqnarray}}
\def\eea{\end{eqnarray}}
\def\bwt{\begin{widetext}}
\def\ewt{\end{widetext}}

\newcommand{\grdint}[2]{\int \text{d}^{#1} #2 \, \sqrt{g} \,}

\def\bnabla{\overline{\nabla}}

\def\bR{\overline{R}}

\def\bg{\overline{g}}

\def\blambda{\overline{\lambda}}

\def\tGamma{\widetilde{\Gamma}}
\def\bGamma{\overline{\Gamma}}

\def\Gammagf{G}

\def\eps{\varepsilon}